\documentstyle[paper,mydefs,12pt,epsf,epsfig]{article}

\newcommand{\fund}{\drawsquare{6.5}{0.4}}

\renewcommand{\Box}{\fund}

\newcommand{\mt}{{M_{T}}}
\newcommand{\mr}{{M_{R}}}
\newcommand{\xx}{{|X|}}
\def\drbar{\overline{\rm DR}}
\def\drbarp{\overline{\rm DR}'}

\def\O{{\cal O}}

\begin{document}
\begin{titlepage}

\vskip-.3in
\preprint{CERN-TH/98-47\\
SLAC-PUB-7764\\
UMDHEP 98-47\\
hep-ph/9803290}

\title{Supersymmetry-Breaking Loops from\\\medskip
Analytic Continuation into Superspace}

\author{Nima Arkani--Hamed$^*$%
\ \ Gian F. Giudice$^{\dagger}$%
\footnote[4]{On leave of absence from INFN, Sez. di Padova, Italy.}%
\ \ Markus A. Luty$^{**}$%
\footnote[3]{Sloan Fellow.}%
\ \ Riccardo Rattazzi$^\dagger$}%

\bigskip
\address{$^*$Stanford Linear Accelerator Center,
Stanford University\\
Stanford, California 94309, USA}

\vskip0.3truecm
\address{$^\dagger$Theory Division, CERN\\
CH-1211, Gen\`eve 23, Switzerland}

\vskip0.3truecm
\address{$^{**}$Department of Physics, University of Maryland\\
College Park, Maryland 20742, USA}

\begin{abstract}
We extend to all orders in perturbation theory a method to calculate
supersymmetry-breaking effects by analytic continuation of the 
renormalization group
into superspace.
A central observation is that the renormalized
gauge coupling can be extended to a real vector superfield,
thereby including soft breaking effects in the gauge sector.
We explain the relation between this vector superfield
coupling and the ``holomorphic" gauge coupling, which is
a chiral superfield running only at 1 loop.
We consider these issues for a number of regulators,
including 
dimensional reduction.
With this method, the 
renormalization group equations
for soft supersymmetry breaking terms are directly 
related to
\susc beta functions and  anomalous dimensions
to all orders in perturbation theory.
However, the real power of the formalism lies 
in computing finite soft breaking effects corresponding to
high-loop component calculations.
We prove that the gaugino mass in gauge-mediated
\susy breaking is ``screened'' from strong interactions in the
messenger sector.
We present the complete next-to-leading  calculation of gaugino masses
(2 loops) and sfermion masses (3 loops) in minimal gauge mediation,
and several other calculations of phenomenological relevance.
\end{abstract}

\end{titlepage}

\section{Introduction}
Recently there has been a great deal of interest in building models in
which supersymmetry breaking is communicated to the observable particles
through renormalizable interactions \cite{GMSB}.
A common feature of these models is that \susy breaking occurs in the
masses of ``messenger'' fields in the form
\beq
M = M_{\rm SUSY} + \de M,
\eeq
where $M_{\rm SUSY}$ is a supersymmetric mass term, and
$\de M$ breaks {\susy}.
In most models of this kind constructed to date $\de M \ll M_{\rm SUSY}$,
 and so the messenger threshold is approximately \susc.
Integrating out the messenger fields gives rise to \susy breaking in
the low-energy effective lagrangian below the scale $M$.
A large amount of work has already been done on the calculation of the
\susy breaking effects from various types of interactions
\cite{GMSBcalc,GR}.
In \Ref{GR} it was shown how to compute the leading low-energy
\susy breaking effects in a large class of models using
only one-loop renormalization group (RG)
equations and tree-level matching,
while direct calculations of the same quantities require the evaluation
of 1- and 2-loop graphs.

The starting point of \Ref{GR} is the observation that
since the messenger threshold is approximately \susc,
one can use a formalism where all couplings and masses are treated
as superfields, and the SUSY breaking terms correspond to non-zero
$\th$-dependent spurion components of the couplings.
In this framework, it is not hard to see that leading-log
effects that are determined by the RG in the
SUSY limit are related to \emph{finite} SUSY-breaking effects.
For example, the RG can be used to compute corrections of the form
$(\ln M) / (16\pi^2)$, where $M$ is a threshold mass.
If $M$ is a superfield, then this contribution has
a SUSY-breaking component
\beq
\frac{1}{16\pi^2} \left. \ln M \right|_{\th^2\bar{\th}^2}
= \frac{1}{16\pi^2} \frac{\left. M \right|_{\th^2\bar{\th}^2}}
{\left. M \right|_0},
\eeq
which contains a loop factor, but no logarithm.
Effects of this type therefore correspond to finite loop effects
that are not related to an RG calculation in components.

A simple power-counting argument can be used to show that
in gauge-mediated models
the leading SUSY-breaking terms in the low-energy effective lagrangian
arise from this sort of threshold dependence in the dimensionless
couplings.
This allows one to compute 1- and 2- loop SUSY breaking effects using
the 1-loop RG equations and tree-level matching,
analytically continued into superspace.
In \Ref{GR} this technique was used to reproduce known results in a much
simpler way, and also to derive new phenomenologically interesting results
that would be much more difficult to compute directly.

In this paper, we extend the analysis of \Ref{GR} to higher orders in
perturbation theory.
One motivation for this is to define an unambiguous procedure to perform the
analytic continuation into superfield beyond one loop.
We show that the gauge coupling 
is naturally
extended to a real superfield that is not
the sum of a chiral and an antichiral superfield.
The $\th^2\bar{\th}^2$ component
of the real gauge superfield plays a crucial role in reproducing the
correct behavior of perturbation theory.
Another motivation for this is to obtain new results of interest for
testing models in the literature.
In particular, we are able to compute gaugino, squark, and slepton
masses in gauge-mediated models at the next-to-leading order in
perturbation theory. Our result corresponds to an explicit calculation
of 2- and 3-loop Feynman diagrams.
One of our results is that the gaugino masses in gauge-mediated
models are ``screened'' from corrections from the SUSY-breaking
sector up to 4 loops.
This implies that the gauge-mediation relations are preserved up to
corrections of order $g_{\rm SM}^4 / (16\pi^2)^2 \sim 10^{-4}$
even if the SUSY-breaking (or messenger) sector is strongly coupled.
We also compute other interesting effects, like
the gaugino masses in ``mediator" models~\cite{Lisa},
the gauge-mediated effective potential
induced along classically flat directions, both for $D$ flat directions (2
loops) as well as for the scalar partner of the axion (3 loops).


This paper is organized as follows.
In Section 2, we give a definition of renormalized coupling constants
that can be viewed as superfield spurions to all orders in perturbation
theory.
We use as examples specific theories that allow 
simple \susc regulators.
In Section 3, we discuss this prescription in the case in which the
theory is regulated using 
dimensional reduction.
We also 
show that
extending the couplings to superfields automatically selects the
so-called $\drbarp$ scheme for the soft terms.
In Section 4,
we use our technique to prove the gaugino screening result mentioned
above, and compute gluino, squark, and slepton masses in gauge mediation
at the NLO.
We also extend our results to $D$-term breaking of SUSY,
and derive the gaugino mass in ``mediator" models.
In Section 5, we compute some other interesting SUSY-breaking
effects in gauge-mediated theories.
Section 6 summarizes our main results and contains our conclusions.

\section{Renormalized Coupling Constants as Superfields}
The main tool of our approach is the use of renormalization schemes in
which the renormalized coupling constants can be
treated as superfields.
Much of our discussion can be viewed as a restatement of the
insights of Shifman and Vainshtein~\cite{Russians}
in the framework of renormalized perturbation theory.
However, we will generalize the method to include \susy breaking effects.
For gaugino masses and $A$ terms, this was first done
in ref.~\cite{hisano}. Here we will simultaneously describe the running
of the scalar masses. For related studies, see also
ref.~\cite{altri}.

\subsection{Invitation: the Wess--Zumino Model}
\label{wesszum}
In this subsection we consider a simple example that illustrates many of
the main ideas we will use in more complicated theories.
We consider a massless Wess--Zumino model with bare lagrangian
\beq
\scr{L}_0 = \myint d^4\th\, \scr{Z}_0 \Phi^\dagger \Phi
+ \left( \myint d^2\th\, \frac{\la}{3!} \Phi^3 + \hc \right),
\eeq
and higher-derivative regulator terms~\cite{ilio}
\beq
\scr{L}_{\rm reg} = \myint d^4\th\, \scr{Z}_0
\Phi^\dagger \frac{\Box}{\La^2} \Phi.
\eeq
We can incorporate soft SUSY breaking by extending the bare couplings
$\la$ and $\scr{Z}_0$ to be $\th$-dependent (but $x$-independent)
superfields.%
\footnote{Note that taking a superfield $S$ to be $x$-independent
does not violate SUSY, since it amounts to imposing the \susc
constraint $\partial_\mu S = 0$.}
($\la$ is a superpotential coupling, and is not renormalized.)
We have regulated the theory in a \susc manner, so we can treat the
bare couplings as superfields even at the quantum level.

Because the theory is regulated in a way that preserves SUSY
(including the spurious SUSY acting on the couplings),
the divergences that appear order-by-order in perturbation
theory can be absorbed by \susc counterterms.
That is, we can write
\beq
\scr{Z}_0 = \scr{Z}(\mu) + \de\scr{Z}(\la, \scr{Z}(\mu), \La/\mu),
\eeq
where $\de\scr{Z}$ is the matter wavefunction counterterm.
Because the relation between the bare and renormalized couplings
preserves SUSY, we see that the renormalized couplings can also be
viewed as SUSY spurions.

More specifically, we can define the counterterms by computing
supergraphs with renormalized couplings in the vertices and propagators
and choosing the counterterms to cancel the divergences.
In the SUSY limit where there is no $\th$ dependence in $\scr{Z}_0$
and $\la$, the counterterms have the form \cite{supergraph}
\beq\eql{WZct}
\de\scr{L} = \myint d^4\th\, \scr{Z}\,
C(|\la|^2 / \scr{Z}^3(\mu), |\La|/\mu)
\Phi^\dagger \Phi,
\eeq
where the form of the function $C$ follows from the fact that theory
depends trivially on the overall normalization of the fields.

In the presence of soft SUSY breaking, the renormalized couplings $\scr{Z}$
and $\la$ will also depend on $\th$, and there are new terms in the
Feynman rules involving supercovariant derivatives acting on the couplings
$\scr{Z}$ and $\la$.
However, it is easy to see that such terms can be ignored for purposes of
computing the counterterms~\cite{yam}.
Because our regulator preserves the spurion SUSY even in
the presence of soft SUSY breaking, we know that the counterterms can
still be chosen to be superfield functions of $\la$ and $\scr{Z}$.
But local superspace counterterms involving supercovariant derivatives of
$\la$ and $\scr{Z}$ are forbidden simply by dimensional analysis.
We conclude that even in the presence of soft SUSY breaking, the counterterms
are still given by \Eq{WZct}.
Note what has happened here:
the renormalization of the theory with soft SUSY breaking is completely
determined by a \emph{\susc} calculation.
This is the advantage of treating the bare and renormalized couplings as
superfields.

The fact that the theory depends in a trivial way on the scale of the fields
can be expressed more formally by noting that the the bare lagrangian is
invariant under
\beq
\Phi \mapsto e^A \Phi,
\quad
\scr{Z}_0 \mapsto e^{-(A + A^\dagger)} \scr{Z}_0,
\quad
\la \mapsto e^{-3A} \la,
\eeq
where $A$ is a $\th$-dependent (but $x$-independent) chiral superfield.
The fact that the relation between the bare and renormalized parameters
preserves this feature can be expressed by stating that the renormalized
parameter $\scr{Z}$ transforms the same way as $\scr{Z}_0$:
\beq
\scr{Z} \mapsto e^{-(A + A^\dagger)} \scr{Z}.
\eeq
If we view this as a $\U1$ ``gauge'' transformation, then
$\ln \scr{Z}$ (and $\ln \scr{Z}_0$) transform as a gauge field.
This point of view will be extremely useful to us later.

The relation between the bare and renormalized quantities determined
by \Eq{WZct}
\beq
\scr{Z}_0 = \scr{Z}(\mu) \left[
1 + C(|\la|^2 / \scr{Z}^3(\mu), |\La|/\mu) \right],
\eeq
determines the RG flow of the theory from $d\scr{Z}_0 / d\mu = 0$.
This gives
\beq\eql{WZgamma}
\mu \frac{d \ln\scr{Z}}{d\mu} = -\mu \frac{d}{d\mu}
C(|\la|^2 / \scr{Z}^3, |\La|/\mu) \equiv \ga(|\la|^3 / \scr{Z}^3).
\eeq
The $\th = \bar{\th} = 0$ component of $\ga$ is just the \susc
anomalous dimension.
The renormalized soft scalar mass $m^2$ is defined by writing
\beq
\scr{Z} = Z\left[ 1 - \th^2 \bar{\th}^2 m^2 \right],
\eeq
where $Z$ is the renormalized wavefunction factor.
The RG equation for the soft mass is determined
by the $\th^2\bar{\th}^2$ component of \Eq{WZgamma}:
\beq
\mu \frac{d m^2}{d \mu}
= -\left. \ga(|\la|^2 / \scr{Z}^3) \right|_{\th^2\bar{\th}^2}
= -\ga'(|\la|^2 / {Z}^3) \frac{3 |\la|^2 m^2}{{Z}^3}.
\eeq
This formula is valid to all orders in perturbation theory.
At the 1-loop level
\beq
\ga = -\frac{1}{16\pi^2} \frac{|\la|^2}{\scr{Z}^3},
\eeq
and we recover the familiar result
\beq
\mu \frac{dm^2}{d\mu} = \frac{3}{16\pi^2}\,
{\bar \la^2 m^2},
\eeq
where $\bar \la= |\la |Z^{-3/2}$ is the running coupling constant.
\Eq{WZgamma} also gives the RG equation
for $A$ terms if we add a non-vanishing $\th^2$
component to $\scr{Z}(\mu)$.

In the following, we will generalize the procedure followed in this
section to general renormalizable SUSY theories with soft SUSY
breaking.
The idea is to include soft SUSY breaking by extending the bare
couplings $K_0$ to $\th$-dependent superfields.
As long as the theory is regulated in a \susc manner, the bare
couplings can be viewed as spurion superfields even at the
quantum level.
We then define renormalized couplings $K(\mu)$ related to the bare
couplings by a superfield relation
\beq\eql{Kdef}
K_0 = G(K(\mu), \La/\mu).
\eeq
The renormalization of the couplings in the SUSY limit then determines
the renormalization of the soft SUSY breaking terms as long as the
relation does not involve supercovariant derivatives acting on $K(\mu)$.
But in a vast class of theories, this is guaranteed by simple
power counting and symmetry arguments.
\Eq{Kdef} therefore determines the complete RG flow of all soft
SUSY breaking parameters.
In the remainder of this Section, we explain how to carry out these
steps for gauge theories, which present additional subtleties.

\subsection{Holomorphic Coupling in Supersymmetric QED}
\label{secqed}
We begin with SUSY QED,
a $\U1$ gauge theory with matter fields $\Phi$ and $\bar{\Phi}$ with
charges $+1$ and $-1$, respectively.
This theory can be regulated in a completely \susc manner using a
combination of Pauli--Villars fields to regulate matter loops and
a higher-derivative regulator for the gauge fields.
The bare lagrangian can be written as
$\scr{L}_0 + \scr{L}_{\rm reg}$,
where
\beq\eql{QEDbare}
\bal
\scr{L}_0 &= \myint d^4\th\,
\scr{Z}_{0} \left( \Phi^\dagger e^{V} \Phi
+ \bar{\Phi}^\dagger e^{-V} \bar{\Phi} \right)
\\
&\qquad +\, \myint d^2\th\, \sfrac{1}{2} S_0 W^\al W_\al + \hc,
\eal\eeq
contains the ``physical'' couplings, and
\beq\eql{regmatter}
\bal
\scr{L}_{\rm reg} &= \myint d^4\th\,
\scr{Z}_{0} \left( \Om^\dagger e^{V} \Om
+ \bar{\Om}^\dagger e^{-V} \bar{\Om} \right)
\\
&\qquad +\, \myint d^2\th\, \La_\Phi \Om \bar{\Om} + \hc
\\
&\qquad +\, \myint d^2\th\, W^\al \frac{\Box}{4\La_G^2} W_\al + \hc,
\eal\eeq
contains the regulator terms.
Here, $\Om$ and $\bar{\Om}$ are Pauli--Villars fields (odd-statistics
chiral superfields) and $\La_\Phi$ and $\La_G$ are cutoffs for the matter
and gauge fields, respectively.
We will take the cutoffs to infinity with $\La_\Phi \sim \La_G$, so there
is effectively a single cutoff.
Note that the bare wavefunction factor $\scr{Z}_0$ appears both in front of
the matter fields and the Pauli--Villars fields.
This is necessary to regularize $\scr{Z}_0$-dependent subdivergences that occur
at two loops and beyond.
For reference, the components of $S_0$ are given by
\beq
S_0 = \frac{1}{2 g_0^2} - \frac{i\Th_0}{16\pi^2}
- \th^2 \frac{m_{\la,0}}{g_0^2},
\eeq
where $\Th_0$ is the (bare) vacuum angle and $m_{\la,0}$ is the
bare gaugino mass.

We incorporate explicit soft SUSY breaking by allowing the bare
coupling $S_0$ and $\scr{Z}_0$ to be superfields
with nonzero $\th$ components.
Just as in the Wess--Zumino model, the fact that the regulator preserves
SUSY means that the bare couplings can be viewed as superfields at the
quantum level, and we can renormalize the theory by adding counterterms
that are local (in superspace) and gauge invariant.
We therefore define renormalized superfield couplings $S$ and $\scr{Z}$
by
\beq
S_0 = S + \de S,
\quad
\scr{Z}_0 = \scr{Z} + \de\scr{Z},
\eeq
where the counterterms $\de S$ and $\de\scr{Z}$ are superfield
functions of $S$ and $\scr{Z}$ determined
order-by-order in perturbation theory to cancel the ultraviolet
divergences.

For $\scr{Z}$ we can proceed exactly as in the Wess--Zumino model
discussed above, but we immediately encounter difficulties when we
try to renormalize the gauge coupling as a superfield.
One way to see the problem is that the only manifestly
gauge-invariant operator that can act as a gauge counterterm is
\beq
\scr{L} = \myint d^2\th\, \sfrac{1}{2} \de S \, W^\al W_\al + \hc
\eeq
However, the result of a supergraph calculation is necessarily a
$d^4\th$ integral.
At one loop, this is not a problem because the one-loop gauge
diagrams are independent of all couplings (since the gauge coupling
is in front of the kinetic term), and the counterterm can be
proportional to
\beq
\myint d^4\th \left( D^\al V W_\al + \hc \right)
= \myint d^2 \th\, W^\al W_\al + \hc
\eeq
However, beyond one loop, the coefficient of the counterterm depends
on the superfield couplings, and the counterterm cannot be written as
$d^4\th$ integral.

This argument can be sharpened by using the fact that the counterterm
$\de{S}$ is a \emph{chiral} superfield.
Because of this, $\de{S}$ must be a holomorphic
function of $S$, $\La_\Phi$, $\La_G$, and $\mu$,
independent of $S^\dagger$ as well as $\scr{Z}$.
We therefore have
\beq\eql{CTtoo}
\de S = f \left( S, \frac{\mu}{\La_\Phi}, \frac{\La_G}{\La_\Phi} \right),
\eeq
where $f$ is a holomorphic function.
Now, the divergence in the gauge coupling $g$ is independent
of the vacuum angle $\Th$ to all orders in perturbation theory,
since $F^{\mu\nu} \tilde{F}_{\mu\nu}$ is a
total derivative, and therefore irrelevant in perturbation theory.%
\footnote{We do not address the subtle question of renormalization
beyond perturbation theory.}
Therefore,
\beq
0 = \frac{\partial \Re(f)}{\partial \Im(S)}
= -\Im \frac{\partial f}{\partial S}.
\eeq
Since $f$ is a holomorphic function, the
only possibility is that $\partial f / \partial S$ is independent of $S$,
which implies
\beq
f(S) = a + b S,
\eeq
where $a$ and $b$ are independent of $S$.
We see that $a$ is the 1-loop contribution, and $b$ is identically zero
(since the zero coupling limit corresponds to $S \to +\infty$).
We conclude that there is no divergence in the vacuum polarization
beyond one loop.%
\footnote{Note that this argument does not assume that $f$ is a power series
in $S$.
This is important
for non-abelian gauge theories, where we will see that
the perturbation series is non-analytic in $S$.}
%
If this argument is to be believed, the coupling $S$ satisfies the
exact (to all orders in perturbation theory) RG equation
\beq
\mu \frac{d S}{d \mu} = -\frac{1}{8\pi^2}.
\eeq
This appears paradoxical, since it is known that the $\beta$-function
has a (scheme-independent) contribution at two loops.

  \begin{figure}
  \hspace{3cm}
  \epsfig{figure=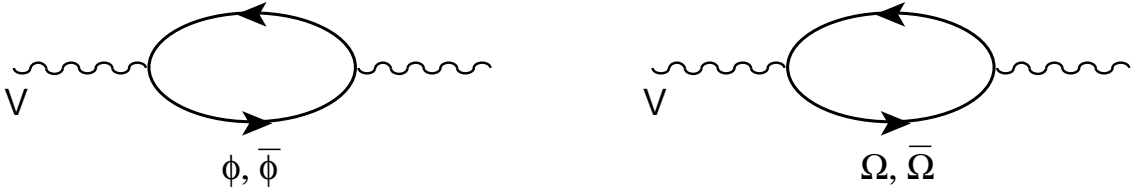,width=15cm}
  \caption[]{\it One-loop diagrams contributing to the vector field
  propagator.
  \label{fig1}}
  \end{figure}

To understand what is going on, we compute the counterterm explicitly
at one loop, keeping the couplings as superfields.
The diagrams are shown in Fig.~1.
We obtain the contribution to the 1PI effective action
\beq
\Ga_{\rm 1PI} = -\frac{1}{2} \myint d^4\th \myint d^4 p\,
V 
\left[\ga(p^2) + \de S + \de S^\dagger \right]
p^2 P_T V + \hbox{finite},
\eeq
where $P_T$ is a transverse superspace projector, and
\beq
\ga(p^2) &= \frac{i}{2} \myint \frac{d^4 k}{(2\pi)^4}\,
\frac{-|\La_\Phi|^2 / \scr{Z}^2}{k^2 (k^2 - |\La_\Phi|^2 / \scr{Z}^2)}\,
\frac{-|\La_\Phi|^2 / \scr{Z}^2}
{(k + p)^2 ((k + p)^2 - |\La_\Phi|^2 / \scr{Z}^2)}
\nonumber\\
&= \frac{1}{8\pi^2} \ln \frac{|\La_\Phi|^2 / \scr{Z}^2}{-p^2}
+ \hbox{finite}.
\eeq
The 1PI effective action can therefore
be made finite by adding the counterterm
\beq\eql{gacoterm}
\de S = -\frac{1}{8\pi^2} \ln\frac{\La_\Phi}{\mu},
\eeq
where $\mu$ is a renormalization scale.
Note that we cannot choose $\de S$ to depend on the ``kinematic'' cutoff
$|\La_\Phi| / \scr{Z}$, the scale at which the Pauli--Villars regulator
cuts off the ultraviolet modes, simply because this quantity is not
a chiral superfield.
On the other hand, it is clear that physical quantities depend on
$\La_\Phi$ only through the combination $|\La_\Phi| / \scr{Z}$,
together with the bare parameters.
This is the key to understanding the meaning of the renormalized
coupling $S$.

More formally, we note that the bare lagrangian is invariant under
\beq\eql{symm}
\bal
\Phi &\mapsto e^{A} \Phi,
\quad
{\bar \Phi} \mapsto e^{A} {\bar \Phi},
\quad
\Omega \mapsto e^{A} \Omega,
\quad
{\bar \Omega}\mapsto e^{A} {\bar \Omega},
\\
\scr{Z}_{0} &\mapsto e^{-(A + A^\dagger)} \scr{Z}_{0},
\quad
\La_\Phi \mapsto e^{-2A} \La_\Phi,
\quad
S_0 \mapsto S_0,
\eal\eeq
where $A$ is a $\th$-dependent (but $x$-independent) chiral
superfield.
Because $\scr{Z}_0$ measures the scale of the fields in the regulated
theory, we can choose the subtractions that defined the renormalized
$\scr{Z}$ so that
\beq
\scr{Z} = \scr{Z}_0 f(S_0 + S_0^\dagger, |\La_\Phi| / \scr{Z}_0,
|\La_G|, \mu),
\eeq
which shows that we can assign the same transformation rule to $\scr{Z}$
as $\scr{Z}_0$.%
\footnote{This may become clearer when we give a 1PI definition of
$\scr{Z}$ in the next Subsection.}
%
%
{}From \Eq{gacoterm} we find that the
renormalized $S(\mu)$ transforms as
\beq\eql{symmtoo}
S \mapsto S - \frac{A}{4\pi^2}.
\eeq
Just as in the case of the Wess--Zumino model, we have found a symmetry
under which $\scr{Z}$ can be interpreted as a background $U(1)$ gauge
field. \Eq{symmtoo} is just a reflection of the Konishi anomaly \cite{Konishi},
therefore  we will refer to this symmetry as the
(renormalized) ``anomalous $U(1)$'' symmetry.%
%
%
As a consequence of this symmetry, physical quantities can depend on $S$
only in the combination
\beq\eql{comb}
S + S^\dagger - \frac{1}{4\pi^2} \ln \scr{Z}
= S_0 + S_0^\dagger + \frac{1}{8\pi^2}
\ln \frac{|\La_\Phi|^2 / \scr{Z}^2}{\mu^2}.
\eeq
(The \rhs shows that this combination depends on the kinematic
cutoff when expressed in terms of the bare parameters.)
%
Notice that $S - S^\dagger$, which is proportional to the vacuum angle,
cannot appear in  any invariant, consistent with the fact that the
vacuum angle is not physical in a theory with massless fermions.
Because of the symmetry defined by \Eq{symm} and \Eq{symmtoo},
the relation between the bare and renormalized wavefunction factors
has the form
\beq
\scr{Z}_0 = \scr{Z}(\mu)~f\left(
S + S^\dagger - \frac{1}{4\pi^2} \ln \scr{Z},
\frac{|\La_{\Phi}| / \scr{Z}}{\mu} ,
\frac{|\La_G|}{\mu} \right).
\eeq
The RG flow of the theory is determined by $d \scr{Z}_0 / d \mu = 0$.
Due to the loop factor multiplying $\ln \scr{Z}$ in the above expression,
an $(n+1)$-loop effects are often related to $n$-loop effects.
There are many examples of this in the literature, and we also obtain
new results of this type in subsequent sections.

Because the correlation functions depend on $S$ only through \Eq{comb},
the relation between the coupling $S$ and a gauge coupling defined directly
in terms of 1PI Green's function is {\it non-analytic} in the couplings.
As already observed in \Refs{Russians},
this can resolve the apparent contradiction between a holomorphic
coupling that runs at one loop and the conventional definition of
the gauge coupling that runs at all loops.

We now note that the quantity
\beq\eql{svztilde}
\tilde{R} \equiv S + S^\dagger - \frac{1}{4\pi^2} \ln\scr{Z}
\eeq
that appears in \Eq{comb} is a good candidate for a \emph{real}
renormalized superfield coupling.
$\tilde{R}$ is a finite quantity that parameterizes the couplings of
the theory, and it does not have any unphysical dependence on the
scale of the fields.
Also, the $\th = 0$ and $\th^2$ components of $R$ give the correct
RG equations for the gauge coupling and gaugino mass to 2-loops.
(In fact, \Eq{svztilde} is identical in form to the famous equation of
\Refs{Russians}, but note our equation involves only renormalized
quantities.)
In the next Subsection, we will 
explain the relation between
$\tilde{R}$ and a renormalized gauge coupling defined
from the 1PI action,
and address the meaning of the $\th^2\bar{\th}^2$ component of
$\tilde{R}$.

We close our discussion of SUSY QED by remarking that
there is a completely analogous $U(1)$ symmetry
with a well defined action on the \emph{bare} couplings.
The ``gauge transformation''
$\Phi \mapsto e^A \Phi$ has an anomaly,
and so the bare gauge coupling must also transform to compensate
for the transformation.
In our regulator, this can be seen from the fact that the Pauli--Villars
fields transform under the symmetry, so the anomaly can be obtained as
the matrix element of the Pauli--Villars mass term in a background gauge
field.
More generally, it is clear that any holomorphic regulator yields the
anomaly,  and the result is that the theory is invariant under the
transformation
\beq
\Phi \mapsto e^A \Phi,
\quad
\scr{Z}_0 \mapsto e^{-(A + A^{\dagger})} \scr{Z}_0,
\quad
S_0 \mapsto S_0 - \frac{A}{4 \pi^2}
\eeq
with the regulator Lagrangian \emph{invariant}.
This ``bare" or ``Wilsonian" anomalous $U(1)$ is also a very
useful symmetry \cite{NimaHitoshi}.

\subsection{Real Superfield Coupling in Supersymmetric QED}
We now give another definition of the renormalized gauge coupling,
obtained directly from the 1PI effective action by subtraction at a
Euclidean momentum point.
This corresponds more closely to the ``physical'' coupling that
describes the momentum dependence of the effective charge.
More to the point, this definition of the gauge coupling
can be directly
understood in terms of component calculations, allowing us
to make contact between our formalism and conventional calculations.

In a component calculation, it is natural to define the renormalized
gauge coupling and gaugino mass in terms of an appropriate 1PI
correlation function at a Euclidean kinematic point.
We now show that a definition of this type gives rise to a real
superfield $R$ whose lowest components are the gauge coupling and
gaugino mass.

Consider the supersymmetric limit first.
To define the renormalized gauge
coupling we must consider the gauge invariant bilinears
in $W_\alpha$ in the 1PI action.
Since we include quantum effects we
must focus on $d^4\th$ integrals.
By a simple operator
analysis one finds there exists just one independent term
\beq\eql{susylim}
\Ga_{\rm 1PI} &= \myint d^4 p
\myint d^4\th \ga(p^2)
W^\alpha \frac{D^2}{-8p^2} W_\al + \hc + \cdots
\\
&= \myint d^4 p \myint d^2\th\,
\sfrac{1}{2} \ga(p^2) W^\al W_\al + \hc + \cdots,
\eeq
where the last identity follows simply by integrating over half of
superspace.
Therefore, $\ga$ can contain the contribution from the
tree-level and loop contributions to the ordinary gauge kinetic term.
We can therefore define the renormalized gauge coupling simply by
subtracting at a Euclidean momentum point:
\beq\eql{gellmann}
\frac{1}{g^2(\mu)} \equiv \left. \ga(p^2) \right|_{p^2 = -\mu^2}.
\eeq
The role of the operator of \Eq{susylim} in generating the
all order $\beta$-function was already emphasized in \Ref{Russians}.

We can similarly define a renormalized wavefunction superfield by
considering the terms in the 1PI action that contribute to the matter
kinetic term
\beq
\Ga_{\rm 1PI} = \myint d^4 p\, \myint d^4\th\,
\zeta(p^2) \left[
\Phi^\dagger e^V \Phi + \bar{\Phi}^\dagger e^{-V} \bar{\Phi} \right]
+ \cdots,
\eeq
and defining
\beq
\scr{Z} = \left. \zeta(p^2) \right|_{p^2 = -\mu^2}.
\eeq

In the presence of soft SUSY-breaking sources in $S$ and $\scr{Z}$,
the gauge kinetic terms in the 1PI effective action are
\beq\eql{softbreak}
\Ga_{\rm 1PI}
= \myint d^4 p \myint d^4\th\,
\ga(p^2) W^\al \frac{D^2}{-8p^2} W_\al + \hc
+ \O(D_\alpha S, D_\alpha \scr{Z}, \ldots)
\eeq
where $\ga(p^2)$ is now a \emph{vector} superfield function of
the couplings $S + S^\dagger$ and $\scr{Z}$,
and $\O(D_\alpha S, \ldots)$
represents terms involving at least one supercovariant derivative
acting on the sources.
By studying all possible $W W$ and $W\bar{W}$ terms involving
supercovariant derivatives, it can be shown that they
always lead to terms of second order in the soft masses, {\it i.e.}
they are $\O(m^2 / p^2)$.
These terms therefore do not contribute to the gauge kinetic term
and gaugino mass term in the 1PI action.
It therefore makes sense to define a renormalized superfield coupling
by
\beq\eql{sgellmann}
R(\mu) \equiv \left. \ga(p^2) \right|_{p^2 = -\mu^2}.
\eeq
Everything in this definition is manifestly \susc, so the relation
between this renormalized coupling and the bare couplings is SUSY
covariant.
The interpretation of the components of $R$ is given by
\beq\bal
\myint d^4\th\, \ga W^\al \frac{D^2}{-8p^2} W_\al
&= \left[ \myint d^2\th\, \sfrac{1}{2} \left( \left. \ga \right|_0
+ \th^2 \left. \ga \right|_{\th^2} \right) W^\al W_\al
+ \hc \right]
\\
&\qquad\qquad
+\, \left. \ga \right|_{\th^2\bar{\th}^2}
\frac{\la^\al \si^\mu_{\al\dot\be}p_\mu \bar{\la}^{\dot\be}}{-p^2}.
\eal\eeq
The lowest components of $R$ are therefore the coefficients of the
gauge kinetic term and gaugino mass term, and we identify
\beq
\frac{1}{g^2(\mu)} \equiv \left. R(\mu) \right|_0,
\quad
-\frac{m_{\la}(\mu)}{g^2(\mu)} \equiv
\left. R(\mu) \right|_{\th^2}. 
\eeq
Note that this renormalization scheme is mass-independent.

The $\th^2\bar{\th}^2$ component of $R$ multiplies a non-local
SUSY-breaking contribution to the 1PI action.
It is instructive to ask what distinguishes this 
$\O(m^2)$
effect from the other $\O(m^2)$ $WW$ and $W\bar{W}$ operators induced
by the terms involving covariant derivatives acting on the couplings.
To do so it is useful to work in components.
Since there are three component fields $A_\mu$, $\la$, and $D$,
there are in general three independent $\O(m^2)$
corrections to the corresponding self-energies: 
\beq\eql{props}
\bal
\Pi^{\mu\nu}_A(p^2)
&= (p^2 g^{\mu\nu} - p^\mu p^\nu) \left(
1 + \frac{\ka_A^2}{p^2} \right),
\\
\Pi_\la(p^2)
&= \sla{p} \left(
1 + \frac{\ka_\la^2}{p^2} \right)
\\
\Pi_D(p^2)
&= \left( 1 + \frac{\ka_D^2}{p^2} \right),
\eal\eeq
where $\ka_{A,\la,D} = \O(m^2)$.
A simple operator analysis shows that the terms 
involving supercovariant derivatives acting on couplings
generate $\O(m^2)$ corrections that always satisfy the supertrace sum
rule $3 \ka_A^2 - 4 \ka_\la^2 + \ka_D^2 = 0$.
On the other hand the $\th^2\bar{\th}^2$ component of
$R$ is associated to a non zero
supertrace
$2 \left. R \right|_{\th^2\bar{\th}^2} = 3 \ka_A^2 - 4 \ka_\la^2 + \ka_D^2$.
If one computes the effect of the dressed self-energies in \Eq{props}
on the matter self-energy, one finds that the only divergent
contribution is proportional to the supertrace.
This simple exercise clarifies why
the $\th^2\bar{\th}^2$ component of $R$, although associated with a
non-local operator, nonetheless enters into the RG flow equations of the
softly broken theory.

We now discuss the relation between the real superfield gauge
coupling discussed here and the holomorphic gauge coupling described
in the previous Subsection.
Since both are perfectly valid parameterizations of the renormalized
gauge coupling,
we can express $R$ in terms of the holomorphic coupling $S$ and $\scr{Z}$.
The coupling $R$ is clearly invariant under the field rescaling \Eq{symm},
so
\beq
R(\mu) = f \left(
S(\mu) + S^\dagger(\mu) - \frac{1}{4\pi^2} \ln \scr{Z}(\mu)
\right).
\eeq
Demanding that the holomorphic and real couplings coincide at tree level
gives
\beq\eql{magic}
R(\mu) = S(\mu) + S^\dagger(\mu) - \frac{1}{4\pi^2} \ln \scr{Z}(\mu)
+\frac{c}{8\pi^2}
+ \O{(S + S^\dagger)^{-1}},
\eeq
where $c$ is a 1-loop scheme-dependent constant.
Notice that this expression automatically gives the correct 2-loop $\beta$
function.
\Eq{magic} is identical to the famous formula of \Ref{Russians} that
relates the 1PI and ``Wilsonian'' gauge couplings.
However, it is important to remember that the coupling $S$ in our
\Eq{magic} is a renormalized coupling constant.

\subsection{Holomorphic Coupling in Supersymmetric Yang--Mills Theory}
We now consider some additional features that arise in
non-abelian gauge theories,
using the example of a pure  SUSY Yang--Mills theory with gauge group
$\SU{N}$.
We regulate this theory in a \susc way by embedding it into a finite
theory with softly broken $\scr{N} = 2$ SUSY.
The additional fields in the regulated theory consist of a chiral
field $\Phi$ in the adjoint \rep (the $\scr{N} = 2$ superpartner of the
$\scr{N} = 1$ gauge multiplet) and $2N$ hypermultiplets consisting
of chiral fields $\Om^J$ and $\bar{\Om}_J$ ($J = 1, \ldots, 2N$)
in the fundamental and antifundamental  \reps, respectively.

The bare lagrangian of the theory (written in $\scr{N} = 1$ superspace) is
\beq\eql{ntwolag}
\bal
\scr{L}_0 &= \myint d^2\th\, S_0 \tr\left[ W^\al W_\al
- \sfrac{1}{4} \bar{D}^2 (e^{-V} \Phi^\dagger e^V) \Phi \right] + \hc
\\
&\qquad +\, \myint d^4\th \left[
\Om_J^\dagger e^V \Om^J
+ \bar{\Om}^{J\dagger} e^{-V^T} \bar{\Om}_J \right]
+ \left( \myint d^2\th\, \sqrt{2} \Om^J \Phi \bar{\Om}_J + \hc \right)
\\
&\qquad +\, \myint d^2\th\, \left[
\La_\Om \Om^J \bar{\Om}_J
+ \La_G \tr(\Phi^2) \right] + \hc
\eal\eeq
The coefficient of the $\Om^J \Phi \bar{\Om}_J$ interaction is fixed by
$\scr{N} = 2$ SUSY.
The $\scr{N} = 2$ SUSY is broken explicitly down to $\scr{N} = 1$
by the $\Phi$ mass term
(the mass term for $\Om^J$ and $\bar{\Om}_J$ is $\scr{N} = 2$ invariant).
$\scr{N} = 2$ theories are finite beyond one loop~\cite{NtwoFinite}.
With our choice of matter, the 1-loop beta function vanishes and therefore,
in the background gauge, there are no divergences.
The parameters $\La_\Om$ and $\La_G$ therefore act as cutoffs for SUSY
Yang--Mills theory, with the fields $\Phi$, $\Om^J$, and $\bar{\Om}_J$
playing the role of regulator fields.
We will eventually take the limit $\La_\Om, \La_G \to \infty$ with
$\La_\Om \sim \La_G$, so that there is effectively a single cutoff.

We now show that the finiteness of this theory persists when
$S_0$ is a chiral superfield with nonzero $\th$ components.
Any divergences in the 1PI effective action
must be local (in $\scr{N} = 1$ superspace) expressions involving the
superfield couplings of the theory.
Because this theory is renormalizable, the divergences must have the
same form as terms in the lagrangian.
There are no divergences when $S_0$ is a number, so any divergences must
be proportional to SUSY-covariant derivatives acting on $S_0$.
But such terms have positive mass dimension, so there can be no
divergences proportional to dimension-4 operators.
The only remaining possibility is that there are divergences proportional
to
\beq
\myint d^2\th\, \bar{D}^2 S_0^\dagger\, \Om^J \bar{\Om}_J + \hc
\quad\hbox{or}\quad
\myint d^2\th\, \bar{D}^2 S_0^\dagger\, \tr(\Phi^2) + \hc
\eeq
Such divergences can be excluded by considering the (anomaly-free) \trans
\beq\bal
\Om^J &\mapsto e^{i\al} \Om^J,
\qquad
\bar{\Om}_J \mapsto e^{i\al} \bar{\Om}_J,
\qquad
\Phi \mapsto e^{-2i\al} \Phi,
\\
&\qquad \La_\Om \mapsto e^{-2i\al} \La_\Om,
\qquad
\La_G \mapsto e^{4i\al} \La_G,
\eal\eeq
under which $\bar{D}^2 S_0^\dagger$ is invariant.

This establishes that the theory above is finite, and therefore provides
a regulator for the SUSY Yang--Mills theory we want to study.
We still need to renormalize the theory in order to take the limit
$\La_\Om, \La_G \to \infty$.
The renormalized lagrangian is%
\footnote{The renormalized lagrangian can be thought of as the
``effective lagrangian'' below the scales $\La_\Om, \La_G$.
However, we must choose the couplings in the ``fundamental lagrangian''
$\scr{L}_0$ as a function of $\La_\Om$ and $\La_G$ so that the couplings in
the ``effective lagrangian'' are held fixed as the cutoff is removed.
This can be thought of as ``fixing the parameters from low-energy
experiment''.}
\beq
\scr{L} = \myint d^2\th\, S \tr( W^\al W_\al) + \hc,
\eeq
where $S$ is defined by
\beq\eql{SctDef}
S_0 = S + \de S.
\eeq
The counterterm $\de S$ is fixed order-by-order in perturbation theory to
cancel the divergences as $\La_\Om, \La_G \to \infty$.

At one loop, the vacuum polarization in the background gauge is proportional to
\beq\eql{SUSYYMvacone}
-\frac{N}{16\pi^2} \ln \frac{|\La_G|^2 / (S + S^\dagger)^2}{-p^2}
- \frac{2N}{16\pi^2} \ln \frac{|\La_\Om|^2}{-p^2}
+ \hbox{finite}
+ \de S + \de S^\dagger,
\eeq
where the ``physical'' cutoff for the $\Phi$ contribution is
$|\La_G| / (S + S^\dagger )$ due to the non-canonical kinetic
term for the gauge multiplet.
At this order, the theory can be renormalized in a holomorphic way
by choosing
\beq\eql{SctEval}
\de S = \frac{N}{16\pi^2} \ln \frac{\La_G}{\mu}
+ \frac{2N}{16\pi^2} \ln \frac{\La_\Om}{\mu},
\eeq
where $\mu$ is a renormalization scale.

Because the theory is regulated in a \susc manner, the same argument
used in Sect.~2.2 shows that there are no counterterms beyond
one loop to all orders in perturbation theory.%
\footnote{Note that the perturbation series is nonanalytic in $S$,
as can be seen from \Eq{SUSYYMvacone}.
However, the arguments of Sect.~2.2 do not require the perturbation
series to be a power series in $S$, and are therefore valid in this
case as well.}
We can therefore choose the counterterm to be given by \Eq{SctEval}
to all orders in perturbation theory.
The renormalized gauge coupling defined in this way
satisfies the \emph{exact} RG equation
\beq
\mu \frac{d S}{d\mu} = \frac{3N}{16\pi^2}.
\eeq

As in SUSY QED, the fact that the holomorphic gauge coupling
has a 1-loop beta function is closely connected to the fact that the
subtraction depends on $\La_\Om$ and $S + S^\dagger$ separately.
Logarithmic divergent loops always involve the ``kinematic" cutoff
$|\La_G| / (S + S^\dagger)$, and therefore the renormalized expansion
coefficient is
\beq
S + S^\dagger + \frac{N}{8\pi^2} \ln(S + S^\dagger)
&= S_0 + S_0^\dagger
- \frac{N}{8\pi^2} \ln\frac{|\La_G| / (S + S^\dagger)}{\mu}
\nonumber\\
\eql{SYMcomb}
&\qquad
- \frac{2N}{8\pi^2} \ln\frac{|\La_\Om|}{\mu}
\eeq

We can also define a real superfield coupling from the 1PI effective
action similarly to what was done for SUSY QED.
In this scheme, there is a real gauge coupling superfield $R$ defined
to be the coefficient of the $V$ propagator term in the 1PI effective
action.
$R$ must depend on the combination \Eq{SYMcomb}, and we find
\beq
R = S + S^\dagger + \frac{N}{8\pi^2} \ln(S + S^\dagger)
+ \O{(S + S^\dagger)^{-1}}.
\eeq

\subsection{General Gauge Theories}
We have so far treated only simple theories where we
know how to construct a manifestly \susc regulator.
However, we now argue that our results apply to any SUSY gauge theory
as long as a \susc regulator exists.
The general arguments above tell us that the only divergence in the
gauge coupling occurs at one loop, and has the form
\beq\eql{genoneloopct}
\de S = \frac{3 T_G}{16\pi^2} \ln\frac{\La_G}{\mu}
- \sum_r \frac{T_r}{16\pi^2} \ln \frac{\La_r}{\mu},
\eeq
where $T_r$ is the Dynkin index of the $r$ representation.
Here $\La_G$ is a cutoff parameter for gauge loops and $\La_r$ is a
cutoff parameter for matter fields in the representation $r$.
Note that in order for this formula to make sense, $\La_G$ and $\La_r$
must be chiral superfield spurions, as they are in the examples
considered previously.
On the other hand, the ``kinematic'' cutoff (the momentum scale at which
loop momenta are damped) cannot be a chiral superfield, for the simple
reason that it must be real.
As we have seen, \Eq{genoneloopct} is consistent with the 2-loop
RG equations provided that the kinematic cutoff for matter loops is
$\La_{r,{\rm kin}} = |\La_r| / Z_r$.
The relation between the kinematic gauge cutoff and $\La_G$ is more
complicated, as seen in the example of SUSY Yang--Mills.
In any case, in order to reproduce the correct 2-loop beta function,
physical quantities must depend on the combination
\beq\eql{gencomb}
R = S + S^\dagger + \frac{T_G}{8\pi^2} \ln(S + S^\dagger)
- \sum_r \frac{T_r}{8\pi^2} \ln Z_r
+ \hbox{2-loop\ corrections},
\eeq
which is the real gauge coupling superfield.
In the following we will give further evidence for the generality of our
conclusions by showing how they arise in dimensional reduction,
a regulator that can in principle be used for any SUSY theory.

\section{Dimensional Reduction}
So far we have been dealing with regulators that apply only to
special theories.
However, in order to be able to calculate higher
order effects in any theory, including the supersymmetric extension
of the Standard Model, the only practical
regulator is dimensional reduction (DRED) \cite{siegel,capper}.
In this section we show how the holomorphic and real gauge couplings
arise in DRED.
We also show that the procedure of analytically continuing the
renormalized couplings into superspace picks out the so-called
${\drbarp}$ scheme \cite{jack} in which the $\ep$-scalar mass does not
appear in physical quantities.

\subsection{Real and Holomorphic Gauge Coupling in Dimensional Reduction}
The renormalization of SUSY gauge theories in the framework of DRED
was clarified more than a decade ago by Grisaru, Milewski and Zanon (GMZ)
\cite{grisaru}.
They pointed out that in $d = 4 - 2\ep$ dimensions, there is an
additional \susc and gauge-invariant local operator
\beq
\scr{O}_{\rm GMZ} = g_{\ep}^{\mu\nu} \tr( \Ga_\mu \Ga_\nu),
\eeq
where $g_\ep^{\mu\nu}$ is the metric in the $2\ep$ ``compactified''
dimensions, and
$\Ga_\mu$ is the superfield gauge connection defined by
\beq
\Ga^\mu = \frac{1}{2} \si^{\mu}_{\al\dot{\be}}
\bar{D}^{\dot\be} \left( e^{-V} D^\al e^{V} \right).
\eeq
This operator is an $\O(\ep)$ (or ``evanescent'') operator, with the
property that
\beq\eql{epskin}
\myint d^4\th\, \scr{O}_{\rm GMZ}
= \ep \myint d^2\th\, \tr( W^\al W_\al) + \hc
\eeq
(Note that $g_\ep^{\mu\nu} \Ga_\mu\Ga_\nu$ is real.)
Therefore, the quantity $\int d^4\th\, \scr{O}_{\rm GMZ}$ is a dimension-4
term that can appear as a counterterm for the gauge kinetic term.

Taking this into account, the bare lagrangian is
\beq
\scr{L}_0 = \left( \myint d^2\th\, S_0 \tr( W^\al W_\al) + \hc \right)
+ \myint d^4\th\, T_0 g_{\ep}^{\mu\nu} \tr( \Ga_\mu \Ga_\nu)
+ \hbox{matter terms}.
\eeq
We can incorporate soft SUSY breaking by extending $S_0$ and $T_0$ to
$\th$-dependent superfield spurions.
Because DRED preserves SUSY, we can treat $S_0$ and $T_0$ as superfields
even at the quantum level.
The meaning of the higher components of $T_0$ is given by
\beq\eql{tmean}
\bal
\myint d^4\th\, T_0 g_{\ep}^{\mu\nu} \tr( \Ga_\mu \Ga_\nu )
&= \ep \left[ \myint d^2\th\, \left(
\left. T_0 \right| + \th^2 \left. T_0 \right|_{\th^2} \right)
\tr(W^\al W_\al) + \hc \right]
\\
& \qquad
+ \left. T_0 \right|_{\th^2\bar{\th}^2} g_{\ep}^{\mu\nu} A_{\mu} A_{\nu}.
\eal\eeq
That is, the lowest components of $T_0$ are contributions
to the gauge coupling and gaugino mass, and the $\th^2 \bar{\th}^2$
component is the $\ep$-scalar mass.

We now renormalize the theory by writing
\beq
S_0 = \mu^{-2\ep} \left( S + \de S \right),
\qquad
T_0 = \mu^{-2\ep} (T + \de T),
\eeq
where $\de S$ and $\de T$ are counterterms that are determined order by
order in perturbation theory to absorb the $1/\ep$ divergences.
Note that we include a finite renormalized value for $T$.
This corresponds to including evanescent effects:
the scalar and $\th^2$ components of $T$ are $\O(\ep)$
contributions to the gauge coupling and gaugino mass, and
the $\th^2\bar{\th}^2$ component of $T$ is a renormalized
$\ep$-scalar mass parameter.
We will return to the significance of these parameters below.
If we compute using supergraphs, all divergences appear in the
1PI effective action in the form $\int d^4\th\, \scr{O} / \ep^n$,
where $\O$ is a local (in superspace) gauge-invariant \susc operator,
so the counterterms can be defined to preserve the SUSY acting on the
coupling constants.

\Ref{grisaru} show that at one loop, the divergences
can be absorbed in $\de S$, but at two loops and higher, all divergences
must be absorbed in $\de T$.
This sheds considerable light on the origin of the 2-loop running
of the gauge coupling, as follows.
At 2 loops (and higher), a $1/\ep^2$ pole in $\de T$ will
appear as a result of subdivergences.
By \Eq{epskin}, this corresponds to a $1/\ep$ pole in the counterterm for
the gauge coupling, which affects the beta function.
The fact that a $1/\ep^2$ pole arises only from subdivergences explains
why the higher-loop contributions to the gauge coupling beta function
are determined by the anomalous dimensions of the matter fields.

New features arise if we include soft SUSY breaking by extending the
couplings to superfields.
At one loop, we find an ultraviolet divergent contribution to the
$\ep$-scalar mass:
\beq\eql{divep}
\de {\tilde m}^2_A = \frac{g^2}{4\pi^2} \frac{1}{\ep} \left[
-T_G |m_{\tilde g}|^2 + \sum_r T_r m^2_r \right].
\eeq
%
%
Although this is a finite effect,
it is known that renormalization of the $\ep$-scalar
interactions is required to preserve unitarity
\cite{thooft,jackjones}.
(Indeed an explicit calculation of Poppitz and Trivedi \cite{poppitz}
shows that {\it infrared} divergences arise at 2-loops if the
$\ep$-scalar mass is not  renormalized.)

To subtract the divergence in the $\ep$-scalar mass in a way
that preserves SUSY acting on the couplings, we must add the 1-loop
counterterm
\beq\eql{drct}
\de T = \frac{1}{8\pi^2} \frac{1}{\ep} \left[ T_G \ln (S + S^\dagger)
- \sum_r T_r \ln \scr{Z}_r \right].
\eeq
The logs ensure that the counterterm for the $\ep$-scalar mass has
the correct dependence on the gauge coupling and is independent
of the wavefunction of the matter fields.
Note that the scalar and $\th^2$ components of $\de T$ give rise to
\emph{finite} contributions to the gauge coupling and gaugino mass.
This restores the dependence of the renormalized gauge coupling
on $\ln \scr{Z}$ and $\ln( S + S^\dagger)$.

We now have all the ingredients we need to define the renormalized
holomorphic and real gauge coupling superfields in DRED.
The holomorphic gauge coupling is defined simply by $S$.
Because $\de S$ contains only 1-loop divergences (and $S_0$ is $\mu$
independent), $S$ runs only at one loop.
On the other hand, because of the subtraction in \Eq{drct},
the components of $S$ do not give the renormalized gauge coupling
and gaugino mass.
Rather, these are given by the lowest components of a superfield $R$,
defined by
\beq\eql{drRdef}
R \equiv S + S^\dagger + \ep T + \de T^{(1)},
\eeq
where $\de T^{(1)}$ is the coefficient of $1/\ep$ in $\de T$.
{}From \Eq{drct}, we see that
the quantities $R$ and $S$ satisfy precisely the relation derived
in the previous Section for other regulators and renormalization
schemes:
\beq
R = S + S^\dagger + \frac{T_G}{8\pi^2} \ln (S + S^\dagger)
- \sum_r \frac{T_r}{8\pi^2} \ln \scr{Z}_r
+ \O((S + S^\dagger)^{-1}).
\eeq
The definition \Eq{drRdef} also shows that physical quantities
must depend on $S$ through $R$, since it is the components of $R$ that
multiply the kinetic terms and gaugino mass terms in the lagrangian.

We need to understand what scheme in component calculations is
picked out by the procedure above.
It is useful to define a bare gauge coupling superfield
\beq\eql{rbare}
R_0\equiv S_0+S_0^\dagger +\ep T_0
\eeq
in terms of which the bare gauge coupling and gaugino mass are
\beq\eql{drbare}
\frac{1}{g_0^2} = \left.
R_0 \right|_0,
\qquad
-\frac{m_{\la,0}}{g_0^2} = \left.
\left( S_0 + \ep T_0 \right) \right|_{\th^2}=\left. R_0\right|_{\th^2}.
\eeq
while the renormalized couplings are (see \Eq{drRdef})
\beq\eql{drren}
\frac{1}{g^2} = \left. R \right|_0,
\qquad
-\frac{m_{\la}}{g^2} = \left. R \right|_{\th^2}.
\eeq
The relation between the bare and renormalized couplings is therefore
determined by the components of
\beq
R_0= \mu^{-2\ep} \left(
R + \frac{\de S^{(1)}}{\ep}
+ \sum_{n = 2}^\infty \frac{\de T^{(n)}}{\ep^{n - 1}} \right),
\eeq
where $\de T^{(n)}$ is the coefficient of $1/\ep^n$ in $\de T$.
We assume that $\de S$ and $\de T$ consist of pure $1/\ep$ poles. This
corresponds to
modified minimal subtraction ($\overline{\hbox{MS}}$) if we
rescale $\mu$ appropriately, writing $\mu = \bar{\mu} \sqrt{e^\ga/4\pi}$
and writing all expressions in terms of $\bar{\mu}$.
\Eqs{drbare} and \eq{drren} then show that $g$ and $m_{\la}$ are
precisely the renormalized couplings in $\drbar$.

When we consider the inclusion of matter with soft scalar masses,
the scheme picked out by the procedure above is identical to
$\drbarp$ \cite{jack}.
To understand the issues involved, note that there appears to be an
extra renormalized parameter in DRED, corresponding to an
$\ep$-scalar mass.
This parameter has a non-trivial RG evolution, and so cannot be
set to zero at all scales.
However, the $\ep$-scalar mass is an evanescent effect, and does not
give rise to an additional parameter at the quantum level.
The way this works is that if we renormalize the theory with an arbitrary
$\ep$-scalar mass parameter, it only appears in physical quantities in
the combination \cite{jack}
\beq\eql{drbarp}
m^2_{r,\overline{\rm DR}}(\mu)
- \frac{g^2_{\overline{\rm DR}}(\mu) C_r}{8\pi^2}
{\tilde m}^2_{{A},\overline{\rm DR}}(\mu)
+ \O(g^4).
\eeq
One can then define the scheme $\drbarp$ by declaring the combination above
to be the renormalized soft scalar mass.
$\drbarp$ is therefore the scheme in which the $\ep$-scalar mass does not
appear in any renormalized expression \emph{for arbitrary values of} $\mu$.

In terms of the superfield couplings, the renormalized $\ep$-scalar
mass corresponds to the term
$\ep \th^2\bar{\th}^2$
 in $R$.
But because we subtract all the $1/\ep$ poles in $R$, the
1PI action is a finite function of $R$.
Therefore, there is no \emph{explicit} dependence on ${\tilde m}_A^2$
in physical quantities, for any value of $\mu$.
This is sufficient to prove that the scheme we have defined is
identical with $\drbarp$.
Our procedure extends the definition of
$\drbarp$, given in ref.~\cite{jack} at the 2-loop level,
to all orders in perturbation theory.

Note that
the inclusion of the evanescent $\ep T$ term
in \eq{drRdef} is essential for $R$ to satisfy the $d$-dimensional
RG equation
\beq\eql{rbeta}
\mu \frac{dR}{d\mu} = 2 \ep R + \beta (R).
\eeq
This is easy to check at lowest order by considering
the RG equation for $T$.
Therefore, in our scheme $\tilde m_A^2$
plays a role similar to that of the $\O(\ep)$ term in the $d$-dimensional
RG equation for $g^2$:
it insures $\mu$ independence of the bare coupling $g_0^2$,
but is irrelevant in calculations.

To see more explicitly the connection to \naive $\drbar$, consider
the relation  between the bare and renormalized wavefunctions for the
matter fields
\beq
\scr{Z}_{r,0} = \scr{Z}_r \left[ 1 + \sum_{n = 1}^{\infty}
\frac{\de \scr{Z}^{(n)}_r(R)}{\ep^n} \right].
\eeq
Taking the $\th^2\bar{\th}^2$ components of both sides gives
\beq\eql{drt}
m^2_{r,0} = m^2_{r} - \frac{d}{dR} \Bigl[
\de \scr{Z}_r^{(1)}(R) \Bigr]
\tilde{ m}_A^2 + \hbox{$1/\ep$\ poles}.
\eeq
In our scheme, the renormalized scalar mass is
$m_r^2 = -\left. \ln \scr{Z} \right|_{\th^2\bar{\th}^2}$, while
the finite term on the \rhs is the scalar mass in $\drbar$
(\emph{not} $\drbarp$), since it corresponds to minimal
subtraction.
Comparing \Eqs{drt} and \eq{drbarp}, we see that $m_r^2$ is identical
to $m^2_{r,\drbarp}$ to 2 loops.
(But note that our scheme is defined to all orders in perturbation
theory.)

Let us summarize the main results.
In the supersymmetric limit where the explicit soft breaking is turned
off, we can renormalize the theory by (modified) minimal subtraction,
defining renormalized couplings in the $\drbar$ scheme.
Our result is that if we include renormalized soft terms by
analytically continuing both the renormalized
couplings and the counterterms (defined as functions of the renormalized
couplings) into superspace via
\beq\eql{continue}
\frac{1}{g^2} \to R,
\qquad
Z_r \to \scr{Z}_r,
\eeq
this defines a valid subtraction scheme for the softly-broken theory.
This picks out a unique scheme for the soft terms to all orders in
perturbation theory, which we call $\overline{\hbox{SDR}}$ for
\susc dimensional reduction.
(At two loops, $\overline{\hbox{SDR}}$ coincides with $\drbarp$, so we can
think of it as an all-orders definition of $\drbarp$.)
In $\overline{\hbox{SDR}}$, the RG equations for all soft parameters is
determined by the RG equations in the SUSY limit, to all orders in
perturbation theory. For instance, in gauge mediated models (see
the next section), the analytic continuation of \Eq{continue} is
simply performed by substituting $M\to M+\theta^2 F$ in the effective couplings
of the low-energy supersymmetric Standard Model.


We close with two comments on the superfield coupling
$R$ defined above.
Note that the {\it finite} $\th^2\bar{\th}^2$ component of $R$ defined in DRED
corresponds to an \emph{infinite} contribution to the $\ep$-scalar
mass.
In our definition of $R$ from the 1PI effective action, the
$\th^2\bar{\th}^2$ component of $R$ was related to a nonlocal
effect.
It is interesting to see the connection between these effects
explicitly by considering softly broken SQED
as in Section 2.2, but dimensionally reduced to
$4 - 2\ep$ dimensions.
After subtracting the $\scr{Z}$ independent $1/\ep$ divergence
the gauge self-energy has the form
\beq\eql{dranomaly}
\bal
\Ga_{\rm 1PI} &= \frac{1}{4\pi^2} \myint d^4\th\, \ln \scr{Z}
\left[
\frac{1}{\ep} g_\ep^{\mu\nu} \tr(\Ga_\mu \Ga_\nu + \hc)
+ \tr \left( W^\al \frac{D^2}{-p^2} W_\al + \hc \right) \right]
\\
& \qquad + \hbox{($\scr{Z}$-independent)}+\O(D_\alpha \scr{Z},\dots).
\eal\eeq
If we write this out in terms of components of $\scr{Z}$, we see
that the the terms involving $\left.\scr{Z}\right|$ and
$\left.\scr{Z}\right|_{\th^2}$
are local and exactly cancel between the two terms in brackets.
What is left, from $\left.\ln \scr{Z}\right|_{\theta^2\bar\theta^2}$,
is just a divergent $\ep$-scalar mass, see \Eq{divep},
and a non-local correction to the gaugino self-energy, see \eq{props}.
Anyway, we must subtract the divergent $\ep$-scalar  with
a superfield counterterm as \eq{drct}, so that in the
subtracted  1PI, the dependence on $\ln \scr{Z}$ is all coming from the
non-local operator.
This shows that the
``chiral'' components of $R$ defined in
DRED and by 1PI subtraction differ only by finite analytic
($\scr{Z}$-independent) terms, that is, by a change in scheme.
In this sense, the two definitions are equivalent.

A closely-related issue involves the relation between the origin of the
$\ln\scr{Z}$ term in $R$ in DRED and in the general discussion of
SQED given earlier, where it was inferred from the anomalous $U(1)$
symmetry.
It is conventionally said that there is no rescaling or chiral
anomaly in DRED, and it may appear that there is no direct connection
between these arguments.
However, an intriguing clue can be seen by considering the bare Lagrangian
with couplings $S_0$, $T_0$, and $\scr{Z}_0$.
This Lagrangian has the symmetry
\beq
T_0 \mapsto T_0 + A + A^\dagger,
\quad
S_0 \mapsto S_0 + \ep A,
\quad
\scr{Z}_0 \mapsto \scr{Z}_0,
\eeq
which ensures that physical quantities depend on the combination
$S_0 + S_0^\dagger + \ep T_0$.
However, arbitrary values of $T_0$ lead to inconsistencies
(loss of unitarity and IR divergences). Up to two loops the choice
\beq
T_0 = -\frac{1}{4\pi^2} \frac{1}{\ep} \ln \scr{Z}_0
\eeq
eliminates the problems.
But with this choice, physical quantities depend on the combination
$S_0 + S_0^\dagger - \ln\scr{Z}_0 / 4\pi^2$, which is just what is
required to obtain the anomalous $U(1)$.
We believe that these are very suggestive connections that come close to
exposing the anomaly in DRED, and we plan on exploring this point more
completely elsewhere.

\subsection{Two-loop Renormalization Group equations in $\drbarp$}
We can check explicitly that the scheme defined above is equivalent to
$\drbarp$ at NLO
by computing the 2-loop RG equations for the gluino and sfermion masses.
Consider the real gauge coupling, given by
\beq\eql{betafield}
R(\mu) = S(\mu) + S^\dagger(\mu)
+ \frac{T_G}{8 \pi^2} \ln\left[
S(\mu) + S^\dagger(\mu) \right]
- {T_r\over 8 \pi^2}\ln \scr{Z}_r(\mu),
\eeq
where $S$ is the holomorphic gauge coupling.
The gaugino mass is given by
$m_{\la}=~-\left.\ln R \right|_{\th^2}$, so its
NLO $\beta$ function is easily derived from \Eq{betafield}:
\beq\eql{gauginonlo}
\mu \frac{d m_\la}{d\mu} = -{g^2 \over (8\pi^2)^2}
\left( T_G b - 2 \sum_r T_r C_r \right) m_\la. 
\eeq
where $b=3T_G-\sum_r T_r$.
This equation agrees with the explicit component calculations in
$\drbar$.
A similar derivation, based on the Konishi anomaly, was given
by Hisano and Shifman~\cite{hisano}.
A new feature of the present treatment is that $R$ also governs the
evolution of the dimension-2 soft terms.
To see this, consider
\beq\eql{str}
\left. R\right|_{\th^2\bar{\th}^2}
= {1\over 8\pi^2} \left[
-T_G m_{\la}^2 + \sum_r T_r m_r^2 \right].
\eeq
According to our discussion above, $\left. R \right|_{\th^2\bar \th^2}$
corresponds to a $1/\ep$ counterterm for the $\ep$-scalar mass.
\Eq{str} 
agrees with what is found in explicit
component calculations \cite{poppitz}.
(Notice that the quantity 
on the \rhs
is proportional to the supertrace weighted by the Dynkin indices.)
Now, consider the 2-loop RG equation for matter fields in $\drbar$
\cite{MaVaold,MaVa}
\beq\eql{wavenum}
\mu {d \ln Z_r\over d \mu} = \frac{1}{8\pi^2} \left\{
2 C_r g^2 +  {g^4\over 8\pi^2} C_r
\left[ 3 T_G - T - 2 C_r \right] \right\},
\eeq
where $T = \sum_r T_r$.
Its continuation into superspace simply amounts to the substitution
$g^2 \to 1/R$, $Z \to \scr{Z}$.
The RG equation for the scalar masses is then obtained
by taking the $\th^2\bar{\th}^2$ component of \Eq{wavenum}.
This gives
\beq\eql{wavesoft}
\bal
\mu \frac{d m_r^2}{d\mu}
&= -\frac{C_r}{8\pi^2} \biggl\{
4 g^2 m_{\la}^2 + \frac{g^4}{8\pi^2} \Bigl[
2 T_G m_{\la}^2 - 2 \sum_s T_s m_s^2
\\
&\qquad\qquad\qquad\qquad\qquad\qquad\quad
+ 6 (3 T_G - T - 2 C_r) m_{\la}^2 \Bigr] \biggr\},
\eal\eeq
which agrees with the result in $\drbarp$ \cite{MaVa,yam,jackjones,jack}.
The same check can be done for the
evolution of $A$- and $B$-terms and in the presence of Yukawa
interactions.

\section{Gauge-mediated Supersymmetry Breaking}
We now show how to apply the formalism of the previous Section to
perform calculations in gauge-mediated SUSY breaking (GMSB) models.
We begin by briefly reviewing the calculation of the leading
gaugino and scalar masses in GMSB, as performed in \Ref{GR}.
We then turn to new calculations at higher loop orders.
The main new result in this Section is that the gaugino masses are
insensitive to the couplings in the messenger sector up to four
loops.
This ``screening theorem'' means that it is possible to make precise
predictions for gaugino masses even when the SUSY breaking dynamics
is strongly coupled.
The scalar masses are not screened in this way, and are therefore
sensitive to strong SUSY-breaking dynamics.
We also compute the NLO corrections to SUSY-breaking masses in GMSB,
which correspond to 2-loop corrections for gaugino masses and
3-loop corrections to the scalar masses.

\subsection{Leading Results}
In this Subsection,
we briefly review the main results of \Ref{GR} for completeness.
Consider the fundamental theory
\beq\eql{gamedla}
\bal
\scr{L}' &= \myint d^4\th \left[
\scr{Z}'_Q \left( Q^\dagger e^{V^{(Q)}} Q
+ \bar{Q}^\dagger e^{V^{(\bar{Q})}} \bar Q \right)
+ \sum_r \scr{Z}'_r q_r^\dagger e^{V^{(r)}} q^{\vphantom{\dagger}}_r \right]
\\
&\qquad
+ \myint d^2\th\, S' \tr( W^\al W_\al ) + \hc
\\
&\qquad
+ \myint d^2\th\, \la X Q \bar{Q} + \hc ,
\eal\eeq
where $Q, \bar{Q}$ are the messengers, $q_r$ are observable sector
fields, and $X$ is a singlet.
$X$ is a background chiral superfield that parameterizes the effect
of SUSY breaking via
\beq
\la X = M + \th^2 F,
\eeq
with the assumption $F \ll M^2$.
Our notation is appropriate to the case where there is a single gauge
group, but our formulas are trivial to generalize to the case of
product gauge groups.
Below the scale $M$, the effective lagrangian is
\beq\eql{stdleff}
\bal
\scr{L} &= \myint d^4\th
\sum_r \scr{Z}_r q_r^\dagger e^{V^{(r)}} q^{\vphantom{\dagger}}_r
\\
&\qquad
+ \myint d^2\th\, S \tr( W^\al W_\al ) + \hc + \cdots,
\eal\eeq
where the omitted terms consist of higher-dimension operators.
The low-energy gauge coupling is given by tree-level matching and
one-loop running to be
\beq\eql{soneone}
S(\mu) = S'(\mu_0) + \frac{b'}{16\pi^2} \ln \frac{M}{\mu_0}
+ \frac{b}{16\pi^2} \ln \frac{\mu}{M},
\eeq
where
\beq
b' = b- N,
\qquad
b = 3 T_G - \sum_r T_r,
\eeq
are the beta function coefficients in the full and effective theories,
respectively. $N\equiv \sum_Q T_Q$ is the ``messenger index".
Here $\mu_0$ is an ultraviolet scale where the theory is defined;
this means that we must
evaluate derivatives holding the running couplings at the
scale $\mu_0$ fixed.

The dependence of the low-energy effective Lagrangian on the
SUSY-breaking effects
is given simply by making the replacement
\beq\eql{analcont}
M \to X
\eeq
in the dependence of the effective couplings $S$ and $\scr{Z}_r$.
(Notice that to simplify the notation we have absorbed $\lambda$ in the
definition of $X$).
It is this ``analytic continuation'' that is at the heart of the method
of \Ref{GR}.
We can now read off the gaugino mass from%
\beq\eql{stdgaugino}
m_\la(\mu) =- g^2(\mu) \left.
\frac{\partial S(\mu)}{\partial X} \right|_0 F
= \frac{N g^2(\mu)}{16\pi^2}\, \frac{F}{M},
\eeq
where the notation ``$|_0$'' denotes setting $\th = \bar{\th} = 0$
and $X = M$.
Note that this automatically gives the correct RG improvement of the
gaugino mass.
\Eq{stdgaugino} involves the holomorphic gauge coupling, which is
equivalent to the real superfield coupling at one loop.
The use of the real gauge coupling is crucial for the higher-order
calculations we do later.

We now consider the contribution to the gaugino mass coming from
higher-dimension operators in the effective lagrangian~\cite{GR}.
Operators in the effective lagrangian consist of analytic
terms in the light fields and the background $X$ and their derivatives
divided by powers of $X$.
The lowest-dimension operator respecting the $\U1_R$ symmetry that can
contribute to the gaugino mass is
\beq\eql{powercorrect}
\de\scr{L} = \frac{c g^2}{16\pi^2} \myint d^4\th \left[
\frac{X^\dagger D^2 X}{| X|^4} \tr(W^\al W_\al)
+ \hc \right].
\eeq
\Eq{powercorrect} gives a contribution to the gaugino mass of order
\beq
\de m_\la \sim m_\la\,
\frac{|F|^2}{|M|^4}.
\eeq
This is negligible if $F \ll M^2$.
It is easy to see that all other higher-dimension operators also
give contributions to the gaugino and scalar masses that are suppressed
by powers of $F^2_X / M^4$.

We now turn to the calculation of the scalar mass, where the correct
continuation into superspace is $M\to \sqrt{XX^\dagger}$ \cite{GR}.
We compute the matter field wavefunction coefficients $\scr{Z}_r$,
whose $\th$-dependence contains SUSY breaking from the dependence
on the threshold at $M$:
\beq\eql{quarto}
m^{2}_r(\mu)
= -\left.
\frac{\partial^2 \ln \scr{Z}_r(\mu)}{\partial X^\dagger \partial X} \right|_0
|F|^2
= -\frac{1}{4}\,
\frac{\partial^2 \ln \scr{Z}_r(\mu)}{(\partial \ln|X|)^2}\,
\left| \frac{F}{M} \right|^2,
\eeq
where we have used the fact that $\ln \scr{Z}_r$ is a vector superfield,
and therefore depends on $X$ through $|X|$.
The 1-loop RG equation for $\scr{Z}_r$ is
\beq
\ga_r = \mu \frac{d \ln \scr{Z}_r(\mu)}{d\mu} = \frac{C_r}{4\pi^2}\,
\frac{1}{S + S^\dagger},
\quad
\ga'_r = \mu \frac{d \ln \scr{Z}'_r(\mu)}{d\mu} = \frac{C_r}{4\pi^2}\,
\frac{1}{S' + S^{\prime \dagger}}.
\eeq
Computing $\scr{Z}_r$ using 1-loop running and tree-level matching, we have
\beq
\ln \scr{Z}_r(\mu) = \int_{\mu_0}^{M}
\frac{d\mu'}{\mu'}\, \ga'_r(\mu')
+ \int_{M}^{\mu} \frac{d\mu'}{\mu'}\, \ga_r(\mu').
\eeq
This gives
\beq
\frac{\partial \ln \scr{Z}_r(\mu)}{\partial \ln\xx}
&= \int_{|X|}^{\mu} \frac{d\mu'}{\mu'}\,
\frac{\partial \ga_r(\mu')}{\partial \ln\xx}
\nonumber\\
&= \frac{C_r}{4\pi^2} \int_{|X|}^{\mu} \frac{d\mu'}{\mu'}\,
\frac{\partial}{\partial \ln \xx}
\left( \frac{1}{S(\mu') + S^\dagger(\mu')} \right).
\eeq
Note that the explicit $\xx$ dependence from the limits of integration
cancels in the derivative because of the tree-level
matching conditions.
{}From \Eq{soneone}, we see that
\beq\eql{vaccam}
S(\mu) + S^\dagger(\mu) = S'(\mu_0) + S'^\dagger(\mu_0)
+ \frac{b'}{16\pi^2} \ln \frac{X^\dagger X}{\mu_0^2}
+ \frac{b}{16\pi^2} \ln \frac{\mu^2}{X^\dagger X},
\eeq
which depends on $X$ only through $|X|$, as required.
We then obtain
\beq
\frac{\partial \ln \scr{Z}_r(\mu)}{\partial \ln \xx}
= -\frac{C_r}{4\pi^2} \int_{X}^{\mu}
\frac{d\mu'}{\mu'}\, \frac{b' - b}{8\pi^2} \left(
\frac{1}{S(\mu') + S^\dagger(\mu')} \right)^2.
\eeq
Computing one more derivative yields
\beq
\left. \frac{\partial^2 \ln \scr{Z}_r(\mu)}{\partial (\ln \xx)^2}
\right|_{\mu =M}
= -\frac{2 C_r N}{(8\pi^2)^2}\, g^4(M),
\eeq
where we used the definition of the messenger index $N \equiv b - b'$.
This gives a scalar mass
\beq
m^2_r (M)
= \frac{C_r N \alpha^2(M)}{8\pi^2}\,
\left| \frac{F}{M} \right|^2.
\eeq
It is remarkable that the finite part of a 2-loop graph can be evaluated
from a 1-loop RG computation.
In the present approach, this arises because $\scr{Z}_r$ depends on $\xx$ only
through the values of running couplings, and derivatives with
respect to $\xx$ therefore bring in extra loop factors.

\subsection{Gaugino Screening}
\label{gauscreen}
We now consider corrections to the gaugino mass.
Very generally, we will find that contributions from messenger
interactions to the
gaugino mass are suppressed by additional loop factors beyond the \naive
expectation, a result we refer to as ``gaugino screening''.
We will see in Sect.~4.4
that the scalar masses are not similarly screened.

The main point is that the holomorphic gauge coupling is given
\emph{exactly} by
\beq\eql{screen}
S(\mu) = \frac{b' - b}{16\pi^2} \ln X
+ (\hbox{$X$-independent}),
\eeq
where $b$ ($b'$) is the beta function coefficient in the effective theory
below (above) the messenger scale.
(If the SM gauge group has a standard embedding
into a larger messenger
group above the messenger scale, then $b'$ is the beta function of the
larger group.)
The physical gaugino mass must be read off from the
$\th$-dependent components of the real superfield gauge coupling.
(As explained above, the holomorphic gauge coupling has unphysical
field rescaling invariance that is not present in physical quantities.)
The real gauge coupling is related to the holomorphic gauge coupling by
\beq\eql{effreal}
\bal
R(\mu) &= S(\mu) + S^{\dagger}(\mu)
+ \frac{T_G}{8\pi^2} \ln\left[ S(\mu) + S^{\dagger}(\mu) \right]
- \sum_{r} \frac{T_r}{8\pi^2} \ln \scr{Z}_{r}
\\
&\qquad
+ \O{(S + S^\dagger)^{-1}}.
\eal\eeq
The dependence on the wavefunction factors $\scr{Z}_r$ contains the information
about the 2-loop RG behavior of the physical couplings.
Since $S$ is just given exactly by \Eq{screen}, and
since the sum over $r$ runs only over the \emph{light} fields,
$R$ is not affected at the NLO by the messenger interactions.
That's all there is to the proof!

Because the leading dependence on the messenger interactions comes
from $\scr{Z}_r$ in \Eq{effreal}, it is easy to see that
\beq\eql{gauginozeta}
\frac{\de m_\la}{m_\la} \sim \left( \frac{g }{4\pi}\right)^4 \,\left[
\frac{g^2_{\rm mess}}{16\pi^2}+\ln \frac{M'}{M} \right] ~.
\eeq
The $(g/4\pi)^4$ factor arises because the messenger fields interact with
matter only at 2 loops. 
The first term in square brackets represents a
threshold correction due to a
messenger coupling $g_{\rm mess}$,
while the term $\ln(M'/M)$ represents
the sensitivity to mass splittings among the messengers.
Such mass splittings will arise if the various messengers have
different Yukawa couplings
$\lambda$ to the same source $X$ (see \Eq{gamedla}).
In the next Subsection, we perform explicit calculations of the
gaugino masses and NLO, and we will see how the screening theorem
manifests itself in detail.
In the remainder of this Subsection, we confine ourselves to some
qualitative remarks.

Consider for example the dependence on the messenger Yukawa coupling $\la$.
At leading order, the low-energy gaugino masses are independent of $\la$,
but
one may naively expect important quantum corrections if $\la$ is large.
This is not an artificial possibility:
if the Yukawa coupling arises from composite
dynamics, the value of $\la$ will be close to the perturbative limit
$\la \sim 4\pi$ at the compositeness scale \cite{SUSYNDA}.
In this case, $g_{\rm mess}^2 / (16\pi^2) \sim 1$ in \Eq{gauginozeta},
but $\de m_\la / m_\la$ is still suppressed by two weak loops.
Therefore, the gauge-mediation gaugino mass relation are
rather insensitive to strong dynamics of the messenger fields even if
$\la$ is close to the perturbative limit.

Another interesting example is the case in which different messenger
fields have different Yukawa couplings to the same \susy breaking
source $X$.
In other words, the various messengers have different masses $M$ 
but the same ratio $F/M$.
For example, in a GUT model with a messenger scale much lower than the
GUT scale, the running of the messenger Yukawa couplings between the
GUT scale and the messenger scale can induce splittings of the
messenger masses of order $(g^2 / (16\pi^2)\ln(M_{GUT}/M)$, which
can be $\O(1)$ even if the messenger Yukawa interactions are unified
at the GUT scale.
Now, \Eq{gauginozeta} shows that, even for $\O(1)$ messenger mass
splittings, the minimal GMSB relation between the different gaugino masses
are only violated by $\O( (g / 4\pi)^4)$.
Therefore, the gaugino masses do not depend on
the assumption of universality of the messenger Yukawa couplings
at the messenger scale even at NLO,
as long as the Yukawa couplings are of the
same order and there is a single source $X$ of SUSY breaking.

Similar considerations apply to models with vector messengers.
In such models, the \vev that breaks SUSY also breaks a larger gauge
group down to the standard-model subgroup.
There are therefore massive gauge bosons charged under the
standard-model gauge group that act as SUSY-breaking messengers.
\Ref{GR} computed the leading contribution of vector
messengers to the scalar and gaugino masses, and showed that the
contribution to the scalar mass-squared is negative.
The leading contribution to the gaugino mass
from the vector messengers also arises at 4 loops, and again has the
order of magnitude given in \Eq{gauginozeta}, where $g_{\rm mess}$
is now the messenger gauge coupling.
This is important because the messenger gauge coupling can be strong
at the messenger scale.
(For example, this occurs in the models of
\Refs{LutyTerningNimaLutyTerning}).

\subsection{Gaugino Masses at the Next-to-Leading Order}
We now compute the NLO corrections to the gaugino masses in $\drbar$.
In components these corrections correspond to threshold
effects at the messenger mass scale described by 2-loop
Feynman graphs, together with the two-loop RG evolution from the messenger
scale to the physical scale.
In our approach,
these corrections can be extracted from the expression of the real
superfield $R$.
In $\drbar$, the NLO matching at the messenger scale
is simply obtained by requiring continuity of $R(\mu)$ at the threshold of
the physical messenger mass \cite{weinberghall}
\beq\eql{mux}
\mu_X^2 =\frac{XX^\dagger }{\scr{Z}_M^2(\mu_X )}.
\eeq
Here $\scr{Z}_M$ is the wavefunction factor for the messenger fields.
Following the notation introduced in sect.~4.1, primed (unprimed)
quantities refer to the theory above (below) the messenger mass scale.
In terms of the
value of $R'$ at an arbitrary high-energy scale $\mu_0$, much larger than
the messenger scale $\mu_X$,
at the low-energy scale $\mu$ we find
\beq\eql{smux}
R(\mu ) &=R'(\mu_X )+\frac{b}{16\pi^2}\ln \frac{\mu^2}{\mu_X^2}
+\frac{T_G}{8\pi^2}\ln \frac{{\rm Re} S(\mu)}{{\rm Re}S(\mu_X)}
-\sum_r \frac{T_r}{8\pi^2}\ln \frac{\scr{Z}_r(\mu)}{\scr{Z}_r(\mu_X)},
\\
R'(\mu_X ) &=
R'(\mu_0 )+\frac{b'}{16\pi^2}\ln \frac{\mu_X^2}{\mu_0^2}
+\frac{T_G}{8\pi^2}\ln \frac{{\rm Re}S'(\mu_X)}{{\rm Re}S'(\mu_0)}
\nonumber \\
\eql{smu}
&\qquad\qquad
-\,\sum_r \frac{T_r}{8\pi^2}\ln \frac{\scr{Z}'_r(\mu_X)}{\scr{Z}'_r(\mu_0)}
-\frac{N}{8\pi^2}\ln \frac{\scr{Z}_M(\mu_X)}{\scr{Z}_M(\mu_0)}~.
\eeq
Here $S(\mu)$ is the gauge coupling at one loop (see \Eq{vaccam}),
and $R'(\mu_0) = \Re S'(\mu_0)$ gives a SUSY-preserving boundary
condition on the gauge coupling.
The sums in the previous equations extend
over the different matter superfields.
Substituting \Eq{smux} into \Eq{smu}, we obtain
\beq
\eql{sfin}
R(\mu) &= R'(\mu_0) + \frac{b}{16\pi^2} \ln \frac{\mu^2}{\mu_0^2}
+ \frac{b' - b}{16\pi^2} \ln\frac{X X^\dagger}{\mu_0^2 \scr{Z}^2_M(\mu_0)}
\nonumber \\
&\qquad +\,\frac{T_G}{8\pi^2}\ln \frac{{\rm Re}S(\mu)}{{\rm Re}S'(\mu_0)}
-\sum_r \frac{T_r}{8\pi^2}\ln \frac{\scr{Z}_r(\mu)}{\scr{Z}'_r(\mu_0)}~.
\eeq
Notice that in this expression the explicit dependence on
$\scr{Z}_M(\mu_X)$ has dropped out.
An implicit dependence appears from higher-order contributions
in the matter wave-function renormalization $\scr{Z}_r(\mu)$.
However, the NLO expression for the gaugino
mass, which requires only the leading contribution to $\scr{Z}_r(\mu)$, is
independent of $\scr{Z}_M(\mu_X)$.
This is 
a manifestation
of the ``gaugino screening" theorem discussed in
Sect.~\ref{gauscreen}.
See see that
at this order in perturbation
theory, the gaugino masses are not affected by new messenger interactions.
Similarly, the $W$-ino and $B$-ino masses
have no $\al_3$ corrections from messenger thresholds, but only from
their RG evolution below the messenger mass.

To obtain the expression of the gaugino mass, we 
take the $F$ component of \Eq{sfin}:
\beq
m_{\la}(\mu) &= -g^2(\mu)R(\mu)|_{\th^2}
\nonumber \\
& = \frac{1}{1 - g^2(\mu) T_G / (8\pi^2)} 
\left\{
\frac{g^2(\mu)}{16\pi^2}
N\frac{F}{M}
+\sum_r \frac{g^2(\mu)}{8\pi^2}T_r
\left. \ln \scr{Z}_r(\mu)\right|_{\th^2} \right\}~.
\eql{piffy}
\eeq
This equation gives the NLO expression of the gaugino mass in terms of
the SUSY-breaking part of the light matter wave functions
$\scr{Z}_r(\mu)$ at the leading order.
To complete the calculation,
we now compute $\left. \ln \scr{Z}_r(\mu) \right|_{\th^2}$
for matter fields including both gauge and Yukawa interactions.
For simplicity, we give the result for a simple gauge group, but the
generalization to a product group is completely straightforward.
The relevant 1-loop RG equations are:
\beq\eql{zwave}
\mu\frac{d}{d\mu} \ln \scr{Z}_r
&=\frac{C_r}{4\pi^2}g^2-\frac{d_r}{8\pi^2} y^2~,
\\
\mu\frac{d}{d\mu} y^2
&=\frac{y^2}{4\pi^2}\left(
\frac{D}{2}y^2-Cg^2\right)~,
\\
\mu\frac{d}{d\mu}g^{-2} &=\frac{b}{8\pi^2}~,
\eeq
where $y$ is the running Yukawa coupling (physically normalized by
appropriate wavefunction factors).
Here, $d_r$ is the number of fields circulating in the Yukawa loop,
and
\beq
C \equiv \sum_r C_r,
\quad
D \equiv \sum_r d_r,
\eeq
with the sum extended to the field participating in the Yukawa interaction.
If $g$ is the QCD coupling, and $y$ is the top-quark Yukawa coupling,
we have
$C=8/3$ and $D=6$.
Taking the $F$ component of the solution of \Eq{zwave} for $\scr{Z}_r(\mu)$,
we obtain the final expression for the
gaugino mass including QCD ($\al_3$) and top-quark
Yukawa ($\al_t= y_t^2/ (4\pi)$) corrections
\beq
m_{\la_J} (\mu) =\frac{\al_J(\mu)}{4\pi}N\frac{F}{M}
&\left[ 1+T_{J}\frac{\al_J(\mu)}{2\pi} +\frac{4\al_3(\mu)}{9\pi}
(\xi -1)\sum_r T_r \right.
\nonumber \\
&\quad
+\,\left. \frac{\al_t(\mu)}{6\pi}
I(\xi)\sum_r T_r d_r\right] ~,
\eeq
where
\beq
\xi=\frac{\al_3(X)}{\al_3(\mu)},
\quad
I(\xi)=1-\frac{16}{7}\xi
+\frac{9}{7}\xi^{16/9}~,
\eeq
where the sum is taken over the colored light fields. This result has been
recently confirmed by an explicit component calculation~\cite{strumia}.
Notice that in the above equations the dependence on the physical messenger
mass appears via $\xi$, and it is of the form given in \Eq{gauginozeta}.

In order to obtain the pole gaugino mass we have to include also
the finite one-loop corrections at the infrared threshold. For the
gluino, in the $\drbar$ scheme they are given by~\cite{polecor}
\beq
m_{\la_3}^{\rm pole}=m_{\la_3}(\mu)\left\{ 1+\frac{3\al_3(\mu)}{4\pi}
\left[ \ln \left(\frac{\mu^2}{m_{\la_3}^2}\right) +{\cal F}\left(
\frac{\tilde m_q^2}{m_{\la_3}^2} \right)\right] \right\} ~,
\eeq
\beq
{\cal F}(x)=1+2x+2x(2-x)\ln x +2(1-x)^2\ln |1-x|~.
\eeq
The function ${\cal F}$ includes the effect of the gluon-gluino and
quark-squark loops in the approximation in which all
squarks have equal mass $\tilde m_q$.
Since we have neglected weak corrections, the $SU(2)\times U(1)$
gaugino masses receive no contributions from infrared thresholds.
The final expressions for the three gaugino masses improved by $\al_3$ and
$\al_t$ corrections are then given by
\beq
m_{\la_3}^{\rm pole}  &= \frac{\al_3(\mu)}{4\pi}N\frac{F}{M}
\left\{ 1+\frac{3\al_3(\mu)}{4\pi} \left[
\ln \left(\frac{\mu^2}{m_{\la_3}^2}\right) +{\cal F}\left(
\frac{\tilde m_q^2}{m_{\la_3}^2} \right) +2+
\frac{32}{9}(\xi -1)\right] \right.
\nonumber \\
&\qquad\qquad\qquad\qquad
+\,\left. \frac{\al_t(\mu)}{3\pi}
I(\xi)\right\}~,
\eql{gluinlo}
\\
m_{\la_2}^{\rm pole} &=\frac{\al_2(\mu)}{4\pi}N\frac{F}{M}
\left[ 1+\frac{2\al_3(\mu)}{\pi} (\xi -1)+\frac{\al_t(\mu)}{2\pi}
I(\xi)\right] ~,
\eql{m2nlo}
\\
m_{\la_1}^{\rm pole} &=\frac{5\al_1(\mu)}{12\pi}N\frac{F}{M}
\left[ 1+\frac{22\al_3(\mu)}{15\pi} (\xi -1)+\frac{13\al_t(\mu)}{30\pi}
I(\xi)\right] ~,
\eql{m1nlo}
\eeq
where $\al_2= \sfrac{5}{3} \al_1$ at the unification scale.

\begin{figure}
\hspace{3cm}
\epsfig{figure=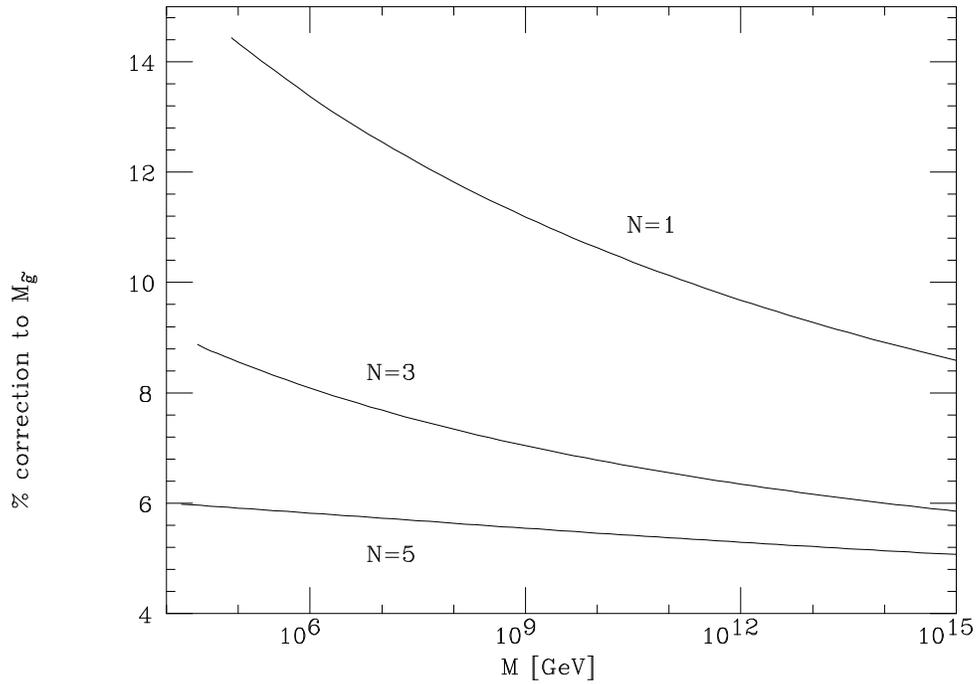,width=15cm}
\vspace{-3cm}
\caption[]{\it NLO correction to the gluino pole mass, as a function of
the messenger mass scale $M$, for messenger index $N=1,3,5$.
We have taken a leading-order value of the gluino mass of 600 GeV and
$\tan\beta=2$,
but the results are rather insensitive to these choices. The curves are
interrupted at values of $M$ that require $F = M^2$ to obtain the required
gluino mass.
\label{fig2}}
\end{figure}

The NLO correction to the gluino mass
$(m_{\la_3}^{\rm pole})^{\rm NLO}/(m_{\la_3})^{\rm LO}-1$ is
shown in Fig.~\ref{fig2}.
We have assumed $(m_{\la_3})^{\rm LO}=600$ GeV and $\tan \beta=2$,
but the result is 
very insensitive to
this choice.
In particular, the value of $\tan \beta$ is 
unimportant
because the top-quark
Yukawa contribution in \Eq{gluinlo} is negligible.
The NLO contribution from messenger loops, which is
obtained by setting $\xi=1$ in \Eq{gluinlo},
is about $+4$--$5\%$.
However, the NLO gauge RG evolution contributes a negative
contribution (see \Eq{gluinlo} and Fig.~2) that almost completely
cancels the messenger contribution for very large running
($M\simeq 10^{15}$~GeV).
The finite gluon-gluino loop gives also a large positive contribution of about
$+10$--$12\%$ to the gluino mass.
This effect is partially compensated by the quark-squark loops,
if the ratio $m_{\tilde{q}}^2/{M}_3^2$ is not large,
as in the case of several messenger flavors ($N>1$).
This explains why the NLO correction to the gluino mass is very important
for small $M$ and $N$, but significantly decreases for larger values of
$M$ and $N$ (see Fig.~2).

The QCD corrections to the $SU(2)\times U(1)$ gaugino masses
vanish at the messenger scale, as expected from the ``screening theorem"
previously discussed.
The effects from the RG running, shown in
\Eqs{m2nlo}--\eq{m1nlo}, tend to cancel
between the gauge and Yukawa term, and give a contribution to the
weak gaugino masses that is at most of few percent.

\subsection{Scalar Masses at the Next-to-Leading Order}
\label{sfermionsec}
We can now also compute the NLO corrections to the squark and slepton
masses in $\drbarp$, which correspond to 3-loop diagrams.
The RG equation for the
wave-function renormalization of 
a matter field $r$ is
\beq
\mu \frac{d}{d\mu} \ln \scr{Z}_r=\ga_r.
\eeq
The gauge contribution to the anomalous dimension $\ga_r$ at the
NLO is given by~\cite{MaVaold,MaVa}
\beq
\ga_r=C_r\frac{g^2}{4\pi^2}+C_r\left[ 3T_G-2C_r- T 
\right] \frac{g^4}
{4(2\pi)^4}~.
\eeq
The SUSY-breaking scalar mass is obtained from \Eq{quarto}
\beq
{\tilde m}^2_r(\mu)&=-\frac{1}{4}\frac{\partial^2 \ln \scr{Z}_r(\mu)}{
(\partial \ln |X|)^2}\left| \frac{F}{M}\right|^2
\nonumber \\
&=
-\frac{1}{4} \left| \frac{F}{M}\right|^2\frac{\partial^2 }{
(\partial \ln |X|)^2}\left[ \int_{\mu_0}^{\mu_X} \frac{d\mu'}{\mu'}\,
\ga'_r(\mu')
\right.
\nonumber \\
&+ \left. \int_{\mu_X}^{\mu} \frac{d\mu'}{\mu'}\,
\ga_r(\mu') \right]~,
\eql{squak1}
\eeq
where $\ga_r$ ($\ga'_r$) is 
the anomalous dimensions below (above)
the physical messenger scale $\mu_X$ (see \Eq{mux}).
Note that $\ga_r$ in the low-energy theory depends implicitly on
$\mu_X$ from the matching conditions at the messenger threshold.
Notice also that the lowest matching correction for the wave function
at the messenger scale $\mu_X$ is at 2-loops. This corresponds
to the addition of an $\O(\alpha(X)^2/16 \pi^2)$ term inside
square brackets in \Eq{squak1}. The resulting correction to the squark mass
is $\O(\alpha^4)$.

For simplicity, we will give the expression of the scalar masses
evaluated at the
messenger scale, as the 2-loop running from $\mu_X$ to the low-energy scale
$\mu$ is well known~\cite{MaVa,yam,jackjones}. In this case, the
action of $\partial^2/
(\partial \ln |X|)^2$ on \Eq{squak1} gives, at the NLO in gauge
interactions,
\beq
{m}^2_r(\mu_X)&=-\frac{1}{4}\left|
\frac{F}{M}\right|^2
\frac{\partial \ln \mu_X}{\partial \ln |X|}
\left. \frac{\partial}{\partial \ln \mu_X}
\left[ \ga_r^\prime(R'(\mu_X))-\ga_r(R(\mu_X))
-\ga_r(R(\mu))\right]
\right|_{\mu=\mu_X} \nonumber \\
&=\frac{g^4}{4}\left|
\frac{F}{M}\right|^2
\left( 1-\ga_M(\mu_X) \right)
\left[ \frac{\partial (\ga_r^\prime(\mu) -\ga_r(\mu))}
{\partial g^2} \frac{\partial
R'(\mu)}{\partial \ln \mu}-\frac{\partial \ga_r}
{\partial g^2} \frac{\partial
R(\mu)}{\partial \ln \mu_X}
 \right]_{\mu=\mu_X}~.
\eql{spree}
\eeq
Here $\ga_M=d\ln \scr{Z}_M /d\ln \mu$
is the anomalous dimension of the messenger superfield
at the leading order,
which depends not only on gauge interactions, but also on any new additional
interactions of the messengers.
In particular, including the Yukawa interaction in \Eq{gamedla}, we
find
\beq
\ga_M=\frac{C_Mg^2}{4\pi^2}-\frac{\lambda^2}{8\pi^2}~.
\eeq
This explicitly shows that the ``screening
theorem", valid for gaugino masses, does not apply to scalar masses.

We can now evaluate the derivatives of $R$, using the expressions obtained
in the previous section:
\beq
\left. \frac{\partial
 R'(\mu_X)}{\partial \ln \mu_X}\right|_0 &=
\frac{b'}{8\pi^2}~,
\eql{rprimo}
\\
\left. \frac{\partial
 R(\mu ,\mu_X)}{\partial \ln \mu_X}\right|_0 &=-
\frac{N}{8\pi^2}\left( 1+\frac{T_G}{8\pi^2}g^2\right) ~.
\eeq
Notice that in \Eq{rprimo} we have kept only the leading term
in the perturbative expansion, since in \Eq{spree} it multiplies
the factor
$\partial (\ga_r^\prime -\ga_r)/\partial g^2$, which is an NLO quantity.
Putting it all together, we obtain the final expression for the scalar
masses at the NLO
\beq
{\tilde m}^2_r(\mu_X) &=\frac{C_r \alpha^2(\mu_X)N}{8\pi^2}\left|
\frac{F}{M}\right|^2 \left( 1-\ga_M \right)
\nonumber \\
&\qquad\qquad
\times\left[ 1+\frac{\alpha (\mu_X)}{2\pi}
\left( T_G-2C_r+N\right)\right] ~.
\eql{squak2}
\eeq

Assuming that
the messengers belong to fundamentals of $SU(5)$,
the NLO expression for the QCD contribution to squark masses is
\beq
{\tilde m}^2_q(\mu_X)=\frac{\al_3^2(\mu_X)N}{6\pi^2}\left|
\frac{F}{M}\right|^2
\left[ 1+\frac{\al_3(\mu_X)}
{2\pi}\left( N-  \frac{7}{3}\right) + \frac{\la_3^2(\mu_X)}{8
\pi^2}\right] ~.
\eql{color}
\eeq
Here $\la_3$ is the messenger Yukawa coupling for the color triplet.
Notice that $\O(\al_3^3 )$ contribution to squark masses from
the messengers tends to cancel the contribution from  gauge and matter
fields, as long as $N$ is not too large.
NLO corrections to slepton masses from QCD and new messenger interactions
come only from the factor $(1-\ga_M )$ in \Eq{spree}.
Since, in our case, weak-doublet messengers are color neutral, the
$SU(2)$ contribution to left-handed slepton masses is corrected only
by the factor $(1+\la_2^2(\mu_X)/8\pi^2)$.  
However, for a generic
choice of messengers, 
the QCD corrections are non-vanishing.
Notice also that in general $\la_2 \ne \la_3$,
although they may be related in a GUT.  
Finally, the improved expression for the right-handed slepton mass is
\beq
{\tilde m}_{e_R}^2(\mu_X)=\frac{5\al_1^2(\mu_X)N}{24\pi^2}\left|
\frac{F}{M}\right|^2
\left[ 1-\frac{8\al_3(\mu_X)}
{15\pi}+ \frac{3\la_2^2(\mu_X)}{40\pi^2}
+ \frac{\la_3^2(\mu_X)}{20\pi^2}
\right] ~.
\eql{slepton}
\eeq
In \Eq{slepton}, $\mu_X$ can correspond to the mass scale of either
the triplet or the doublet messenger mass. The difference between the
two definitions is $\O(\al_1^3 )$, which is negligible in our
approximation.%
\footnote{Higher orders in the electroweak couplings
can be computed following the same procedure
used to obtain \Eq{squak2}, 
with the introduction of separate messenger thresholds.
For an application of the method of \Ref{GR} to the case of
multiple messenger thresholds,
see \Ref{wagner}.}
On the other hand, $\mu_X$ in \Eq{color} has to
be interpreted as the triplet messenger mass, since we include terms
$\O(\al_3^3 )$.

In conclusion, because of the absence of a ``screening theorem", the NLO
corrections to scalar masses are quite dependent on the model assumptions.
They are sensitive to new messenger interactions, like the 
messenger Yukawa
couplings, and they depend on the messenger representations in a way
that cannot be described only by the messenger index $N$.

\subsection{$D$-type Supersymmetry Breaking}
We now consider leading SUSY breaking effects in theories
where the dominant source of SUSY breaking is a $D$-type soft
mass for the messengers rather than a
$F$-type mass, as considered previously.
Some of these results have already been derived in the language of
renormalized couplings in Sect.~3.2.
We discuss them here in a manifestly ``Wilsonian" picture, that is,
by simply computing in the theory with given bare parameters. 
We do this in part for
variety, and in part to show how these results follow from the
``Wilsonian'' anomalous $U(1)$ symmetry.%
\footnote{This symmetry is extremely useful in
obtaining physically interesting results
for non-holomorphic soft terms in strongly coupled SUSY gauge theories
with small soft breakings \cite{RiccNima}.}
Consider a gauge theory with bare lagrangian
\beq
\scr{L}_0 = \myint d^2\th\, S_0 \tr(W^\al W_\al) + \hc
+ \myint d^4\th\, \scr{Z}_{r,0} \Phi_r^\dagger e^{V^{(r)}} \Phi_r,
\eeq
regulated in a \susc manner.
Assume that the theory contains bare soft masses, parameterized by
\beq
\scr{Z}_{r,0} = Z_{r,0} \left[ 1 - \th^2 \bar{\th}^2 m_{r,0}^2 \right].
\eeq
As discussed above, this theory is invariant under the
``Wilsonian'' anomalous $U(1)$ transformation
\beq\eql{wilone}
\Phi_r \mapsto e^{A_r} \Phi_r,
\quad
\scr{Z}_{r,0} \mapsto e^{-(A_r + A_r^\dagger)} \scr{Z}_{r,0},
\quad
S_0 \mapsto S_0 + \sum_r \frac{T_r}{8\pi^2} A_r,
\eeq
with the regulator held fixed.

%

At one loop, the matter terms in the 1PI effective action are
\beq
\Ga_{\rm 1PI} = \myint d^4 p \myint d^4\th\,
\zeta(p^2)
\Phi_r^\dagger e^{V^{(r)}} \Phi_r
+ \hbox{finite},
\eeq
where
\beq\eql{woneloop}
\zeta(p^2) = \scr{Z}_{r,0}
\left[ 1 - \frac{1}{4\pi^2} \frac{C_r}{S_0 + S_0^\dagger}
\ln\frac{\La}{\mu} \right].
\eeq
Here, $\La$ is the ultraviolet cutoff.
Invariance under the transformation \Eq{wilone} allows us to conclude
that $\Ga_{\rm 1PI}$ depends on $S_0 + S_0^\dagger$ only in the
invariant combination
\beq
S_0 + S_0^\dagger - \sum_r \frac{T_r}{8\pi^2} \ln \scr{Z}_{r,0}.
\eeq
This allows us to infer the 2-loop dependence of the matter kinetic
term in $\Ga_{\rm 1PI}$ from \Eq{woneloop}.
We can then obtain
\beq
m_r^2(\mu)
&= -\left. \ln\zeta(p^2 = -\mu^2) \right|_{\th^2\bar{\th}^2}
\nonumber\\
&= m_{r,0}^2 - \frac{g_0^4 C_r}{32\pi^4}
\left( \sum_r T_r m_{r,0}^2 \right) \ln\frac{\La}{\mu}.
\eeq
{}From this, we can read off the 2-loop RG equation for the soft masses
arising from gauge interaction with other soft masses:
\beq
\mu \frac{d m_r^2}{d\mu} = \frac{2 g^4 C_r}{(8\pi^2)^2}
\sum_r T_r m_r^2.
\eeq
(Note we have not specified a definition for the renormalized gauge
coupling, but the result is invariant under changes of scheme for
the gauge coupling.)

%

If the gauge group contains a $U(1)$ factor, there is an additional
contribution to the RG equation for the scalars
from an induced Fayet--Illiopoulos term.
In superspace, a Fayet--Illiopoulos term can be viewed as a
``kinetic mixing'' between the $U(1)$ gauge field and that of the
anomalous $U(1)$ symmetries for the various matter fields.
Note that in the presence of bare soft masses, there is no symmetry
forbidding such a term, so we have an addition contribution
to the bare lagrangian
\beq\eql{fi}
\de\scr{L}_0 = \myint d^2\th\, \sfrac{1}{2} \ka_{r,0} W_1
\scr{W}_{r,0} + \hc,
\eeq
where $W_1$ is the $U(1)$ gauge field strength and
\beq
(\scr{W}_{r,0})_\al
\equiv -\frac{1}{4} \bar{D}^2 D_\al \ln\scr{Z}_{r,0}
= \th_\al m_{r,0}^2
\eeq
is the field strength of the anomalous $U(1)$.
\Eq{fi} contains a linear term in the $U(1)$ auxiliary gauge field
$D_1$, forcing $\avg{D_1} \ne 0$ and giving an additional contribution
to the scalar mass.
It is the running of this contribution that we now compute.

The Fayet--Illiopoulos term is renormalized at one loop, and we obtain
\beq
\Ga_{\rm 1PI} = \myint d^2\th\, \sfrac{1}{2} \left(
\ka_{r,0} + \frac{q_r}{16\pi^2} \ln\frac{\La / \scr{Z}_0}{\mu}
\right) W_1 \scr{W}_{r,0}
+ \hc + \hbox{finite}.
\eeq
Combining this result with the 1-loop renormalization of the matter
wavefunction given in \Eq{woneloop}, we obtain an induced \vev
\beq
\avg{D} = -\frac{g_1^2}{16\pi^2} \left(
\sum_r q_r m^2_{r,0} + \sum_{J,r} \frac{g_J^2}{4\pi^2}
C_r^J q_r m^2_{r,0} \right)
\ln\frac{\La}{\mu},
\eeq
where the sum on $J$ runs over the factors of the gauge group,
and $q_r$ is the $U(1)$ charge of the field $r$.
{}From this we can read off an additional contribution to the RG
equation for the soft mass:
\beq
\mu \left. \frac{d m_r^2}{d\mu} \right|_{D}
= \frac{g_1^2}{16\pi^2} \left( \sum_r q_r m_r^2
+ \sum_{J,r} \frac{g_J^2}{4\pi^2} C_r^J q_r m^2_r \right).
\eeq

Recall that in Sect.~3 we showed that the RG equations for the soft
masses above correspond to $\drbarp$.
The present derivation shows that these RG equations follow as long
as the theory is regulated and subtracted in a \susc fashion.
To further amplify this point, we give an illustrative application
of these methods where we compute a soft mass as a finite calculable
effect.

\newcommand{\Str}{\mathop{\rm Str}}

Consider a toy model with bare lagrangian
\beq\bal
\scr{L}_0 = &\myint d^2 \th\, S_0 \tr(W^\al W_\al) + \hc
\\
&\quad
+\, \myint d^4\th\,
\left[ \scr{Z}_{r,0} \left(
Q_r^\dagger e^{V^{(r)}} Q_r
+ \bar{Q}_r^\dagger e^{V^{(\bar{r})}} \bar{Q}_r \right) \right.
\\
&\qquad\qquad\qquad\quad \left.
+\, \scr{Z}_{q,0}
q^\dagger e^{V^{(q)}} q \right]
\\
&\quad
+\, \myint d^2\th\,
M_r Q_r \bar{Q}_r + \hc,
\eal\eeq
where $r = 1,2$ are two copies of the same gauge representation.
Suppose that the messengers $Q_{1,2}$ have bare soft masses given by
\beq
\scr{Z}_{1,0} = Z_{Q,0} \left[ 1 - \th^2\bar{\th}^2 m_0^2\right ],
\quad
\scr{Z}_{2,0} = Z_{\bar{q},0} \left[ 1 + \th^2\bar{\th}^2 m_0^2\right ].
\eeq
With this choice, the full theory has $\Str \scr{M}^2 = 0$, where
$\scr{M}$ is the full mass matrix of the fields in the theory.
However, if $M_1 \ne M_2$, the effective theory below the scale $M_1$
has nonvanishing mass supertrace.
The value of this supertrace is therefore a calculable
effect in this theory.

We could use the RG equations derived above to compute the soft
masses in the low energy theory.
We present here an alternative derivation of the supertrace
that clarifies the methods used above.
We assume $M_2 \ll M_1$, and compute the $q$ soft mass
in the low-energy theory below the scale $M_2$.
With the choice of parameters made above, we can write
\beq
\scr{Z}_{1,0} = e^{U_0},
\quad
\scr{Z}_{2,0} = e^{-U_0},
\quad
U_0 = -\th^2 \bar{\th}^2 m_0^2.
\eeq
We can view $U_0$ as a ``gauge'' field for a single $U(1)$ under which
$Q_1$ and $\bar{Q}_1$ have charge $+1$, $Q_2$ and $\bar{Q}_2$ have charge
$-1$, $M_1$ has charge $-2$, and $M_2$ has charge $+2$.
Moreover, this $U(1)$ symmetry is \emph{anomaly free}, so we do not
have to appeal directly to a Wilsonian picture of the anomaly.

We now integrate out $Q$ and construct the effective lagrangian below
the scale $M_2$.
This has the form
\beq
\scr{L}'' = \myint d^4\th\,
\scr{Z}''_q q^\dagger e^{V^{(q)}} q
+ \hbox{gauge\ terms},
\eeq
where the $U(1)$ symmetry enforces
\beq
\scr{Z}''_q = f(|M_1| e^{-U_0}, |M_2| e^{U_0}).
\eeq
We can determine the function $f$ by 1-loop running and tree-level
matching to a scale $\mu \ll M_2$:
\beq
\ln \scr{Z}_q(\mu) = \ln\scr{Z}_{q,0}
+ \frac{2 C_q}{b} \ln\frac{g_0^2}{g^2(|M_1|)}
+ \frac{2 C_q}{b'} \ln\frac{g^2(|M_1|)}{g^2(|M_2|)}
+ \frac{2 C_q}{b''} \ln\frac{g^2(|M_2|)}{\mu}.
\eeq
where $b$ ($b'$) [$b''$] are the beta function coefficients in the
full theory (effective theory below $M_1$) [effective theory below $M_2$].
Using the 1-loop expressions for the gauge coupling $g$, and making the
substitution
$|M_1| \to |M_1| e^{-U_0}$,
$|M_2| \to |M_2| e^{U_0}$,
we obtain
\beq\bal
m_q^2(\mu) = -\frac{C_q m_0^2}{4\pi^2} &\left\{
\frac{b - b'}{b''} \left[ g^2(|M_2|) - g^2(|M_1|) \right] \right.
\\
&\qquad
+\, \left.
\frac{b - 2b' + b''}{b''} \left[ g^2(\mu) - g^2(|M_2|) \right]
\right\}.
\eal\eeq
The first term corresponds precisely to the running of the soft mass
between the scales $M_1$ and $M_2$, and the second term to the running
between $M_2$ and $\mu$.
There is no contribution from above the scale $M_1$ because the
contributions from the two messengers cancel.

\subsection{``Mediator'' Models}
We now consider GMSB models where SUSY breaking is
communicated less directly to the observable sector.
We find that 
very generally in such models,
the gaugino screening mechanism
described in Sect.~\ref{gauscreen}
implies the gaugino mass is suppressed compared to the scalar masses
by more loop factors than suggested by a \naive analysis.

We consider the ``mediator'' models introduced in \Ref{Lisa}.
We suppose that a SUSY-breaking sector communicates SUSY breaking
to vectorlike fields $Q$ and $\bar{Q}$.
The fields $Q$ and $\bar{Q}$ are not charged under the
standard-model gauge group.
Rather, they are in a vector-like \rep of a ``mediator'' gauge group
$G_{\rm med}$.
The connection to the observable sector is made through a vectorlike
pair of fields $T$ and $\bar{T}$ that are charged under both the
standard-model gauge group and $G_{\rm med}$.
These fields have a \susc mass term $M_T$ in the lagrangian, which
may be the result of a dynamical mechanism \cite{Lisa}.
The lagrangian of this theory is
\beq\eql{LisaL}
\bal
\scr{L}'' &= \myint d^4\th \biggl[
\scr{Z}''_Q \left(Q^\dagger e^{V_{\rm med}^{(Q)}} Q
+ \bar{Q}^\dagger e^{V_{\rm med}^{(\bar{Q})}} \bar{Q} \right)
+ \sum_r \scr{Z}''_r q_r^\dagger e^{V_{\rm SM}^{(r)}} q_r
\\
&\qquad\qquad
+ \scr{Z}''_T \left( T^\dagger e^{V_{\rm med}^{(T)}} e^{V_{\rm SM}^{(T)}} T
+ \bar{T}^\dagger e^{V_{\rm med}^{(\bar{T})}}
e^{V_{\rm SM}^{(\bar{T})}} \bar{T} \right) \biggr]
\\
&\qquad + \myint d^2\th \left[
M_T T \bar{T}
+ S''_{\rm med} \tr( W_{\rm med}^2 )
+ S''_{\rm SM} \tr( W_{\rm SM}^2 ) \right] + \hc
\\
&\qquad + \de\scr{L}(Q, \bar{Q}, \ldots),
\eal\eeq
where $\de\scr{L}$ contains the interactions that break SUSY.

The holomorphic standard-model gauge coupling below the messenger scales $M$
and the scale $M_T$ is given exactly by
\beq
S_{\rm SM}(\mu) = S''_{\rm SM}(\mu_0)
+ \frac{b''_{\rm SM}}{16\pi^2} \ln\frac{M_T}{\mu_0}
+ \frac{b_{\rm SM}}{16\pi^2} \ln\frac{\mu}{M_T},
\eeq
where $b''_{\rm SM}$ and $b_{\rm SM}$ are the standard-model
beta function coefficients in the effective
theory with and without the field $T$, respectively.
This is independent of $M$, so the leading contribution to the gaugino
mass comes from the $\ln \scr{Z}_r$ term in the real effective gauge
coupling $R_{\rm SM}$, see \Eq{effreal}.
The leading $M$-dependent contribution to $\scr{Z}_r$ arises at 4 loops, so
the gaugino mass arises at 5 loops in this model, as opposed to
the estimate of ref.~\cite{Lisa}.
Since scalar mass-squared terms arise at 4 loops, the gaugino mass is
suppressed compared to the scalar masses in this model, posing a
fine-tuning problem.

To make this argument concrete, and to illustrate the power of
our techniques,
we explicitly compute the gaugino mass in the case
where SUSY breaking is communicated to the fields $q$ and $\bar{q}$
by the \vev of a singlet field $X$:
\beq
\de\scr{L} = \myint d^2\th\, \la X Q \bar{Q} + \hc,
\eeq
with $\avg{X}, \avg{F_X} \ne 0$.
The reader uninterested in details can skip the remainder of this
Subsection.

We will do the calculation for the case where
\beq
M = \la \avg{X} \gg M_T.
\eeq
We further assume that $G_{\rm med}$ is weakly coupled and unbroken
down to the scale $M_T$.
Below the scale $M$, the light fields are $T$, $X$, $Q_r$, $V_{\rm med}$,
and $V_{\rm SM}$, and the effective lagrangian $\scr{L}'$ consists
of the terms in \Eq{LisaL} that depend on these fields.
Below the scale $M_T$, the only light fields are $X$, $Q_r$, $V_{\rm med}$,
and $V_{\rm SM}$, and we denote the effective lagrangian by $\scr{L}$.

Both the scalar and gaugino masses can be read off from $\scr{Z}_r$,
the wavefunction renormalization factor in the low-energy theory.
We therefore compute
\beq
\ln \scr{Z}_r(\mu)
= \int_{\mu_0}^{M} \frac{d\mu'}{\mu'}\, \ga''_r(\mu')
+ \int_{M}^{\mt} \frac{d\mu'}{\mu'}\, \ga'_r(\mu')
+ \int_{M_T}^{\mu} \frac{d\mu'}{\mu'}\, \ga_r(\mu'),
\eeq
where $\ga_r$ ($\ga'_r$) [$\ga''_r$]
denotes the anomalous dimension in the theory $\scr{L}$
($\scr{L}'$) [$\scr{L}''$];
and $M$ ($\mt$) is the matching scale at
the mass of $q$ ($T$), defined similarly to \Eq{mux}.
For example,
\beq
\ga'_r = \mu \frac{d \ln \scr{Z}'_r}{d\mu} = \frac{C_r}{4\pi^2}
\left[ S'_{\rm SM} + S'^\dagger_{\rm SM} - \frac{2T_T}{8\pi^2} \ln \scr{Z}'_T
+ \cdots \right]^{-1},
\eeq
where we have displayed the dependence on $\scr{Z}'_T$ required by the
``anomalous $\U1$'' invariance.
This is important, because $\scr{Z}'_T$ depends on $M$ at 2 loops, giving the
leading $M$ dependence of the anomalous dimensions.
We have
\beq
\frac{\partial \ln \scr{Z}_r(\mu)}{\partial \ln \xx}
= \int_{M}^{\mt} \frac{d\mu'}{\mu'}\,
\frac{\partial \ga'_r(\mu')}{\partial\ln\xx}
+ \int_{\mt}^{\mu} \frac{d\mu'}{\mu'}\,
\frac{\partial \ga_r(\mu')}{\partial\ln\xx},
\eeq
where
\beq
\frac{\partial \ga'_r}{\partial\ln\xx}
= \frac{4 C_r T_T}{(8\pi^2)^2}\,
\frac{1}{(S'_{\rm SM} + S_{\rm SM}^{\prime \dagger})^2}\,
\frac{\partial \ln \scr{Z}'_T}{\partial\ln\xx}.
\eeq
($T$ is not a light field in $\scr{L}$, so there is no contribution
from scales below $M_T$.)
We therefore have
\beq
\frac{\partial \ln \scr{Z}_r(\mu)}{\partial\ln\xx}
= \frac{4 C_r T_T}{(8\pi^2)^2}
\int_{M}^{\mt} \frac{d\mu'}{\mu'}\,
\frac{1}{(S'_{\rm SM}(\mu') + S_{\rm SM}^{\prime \dagger}(\mu'))^2}\,
\frac{\partial \ln \scr{Z}'_T(\mu')}{\partial\ln\xx}.
\eeq
(We see that the $M$-dependent part of $\scr{Z}_r$ is independent of the
renormalization scale $\mu$.)
The dependence of $\scr{Z}'_T$ on the messenger threshold is identical to
the calculation in GMSB, and we obtain
\beq\eql{ddlogzlisa}
\frac{\partial \ln \scr{Z}_r(\mu)}{\partial\ln\xx}
= \frac{8 C_r C_T T_T^2}{(8\pi^2)^4}
\int_{\mt}^{M} \frac{d\mu}{\mu}\,
g'^4_{\rm SM}(\mu)
\int_\mu^{M} \frac{d\mu'}{\mu'}\,
g'^4_{\rm mess}(\mu').
\eeq
{}From this, we can obtain the
gaugino mass
\beq
m_\la(\mu)
&= \frac{g^2_{\rm SM}(\mu)}{2}\,
\frac{\avg{F_X}}{\avg{X}}
\sum_r \frac{T_r}{8\pi^2}\,
\frac{\partial \ln \scr{Z}_r(\mu)}{\partial\ln\xx}
\nonumber\\
&= \frac{4 C_r C_T T_T^2 \left[ \sum_r T_r \right]}{(8\pi^2)^5}
\, g^2_{\rm SM}(\mu)
\nonumber\\
&\qquad \times
\int_{\mt}^{M} \frac{d\mu'}{\mu'}\,
g'^4_{\rm SM}(\mu')
\int_\mu^{M} \frac{d\mu''}{\mu''}\,
g'^4_{\rm mess}(\mu'').
\eeq
Notice that the result scales like $m_\la = \al_{\rm SM}^3
\al_{\rm mess}^2 \ln^2 M/M_T$, indicating that two loops are accounted for
by 1-loop evolution.


\section{Effects from Other Thresholds}
Up to now, we have been focusing on effects that can
be computed from the dependence on the messenger threshold.
However, there are interesting models with other thresholds
than can give rise to important
SUSY-breaking effects in the low-energy theory.
In this Section we analyze some illustrative examples.

\subsection{Flat Direction Effective Potential}
In the limit where SUSY is unbroken,
the minimal \susc standard model has a large space
of flat directions, directions in field space where the classical potential
vanishes identically.
(For an exhaustive list, see \Ref{MSSMflat}.)
All of these flat directions will be lifted by SUSY breaking, and we
are interested in computing the effective potential far out along one
of these flat directions.
For GMSB, the effective potential can be evaluated
from 2-loop component diagrams such as those
evaluated in \Ref{HitoshiFlat}, with the motivation of
studying the cosmology of these flat directions.
We will show how to compute the effective potential without evaluating loop
diagrams.

We will explain our technique using a toy theory with an ``observable
sector'' consisting of a $\U1$ gauge theory with $N_q$ pairs of chiral
fields $q$ and $\bar{q}$ with charges $+1$ and $-1$, respectively.
These are coupled to a ``messenger sector'' consisting of $N_Q$ pairs of
chiral fields $Q$ and $\bar{Q}$ and a singlet field $X$ that parameterizes
SUSY breaking.
The lagrangian is
\beq\bal
\scr{L}'' &= \myint d^4\th\, \Bigl[
\scr{Z}''_q \left( q^\dagger e^V q + \bar{q}^\dagger e^{-V} \bar{q} \right)
\\
&\qquad\qquad
+ \scr{Z}''_Q \left( Q^\dagger e^V Q + \bar{Q}^\dagger e^{-V} \bar{Q} \right)
+ \scr{Z}''_X X^\dagger X \Bigr]
\\
&\qquad + \myint d^2\th\, \sfrac{1}{2} S'' W^\al W_\al + \hc
\\
&\qquad + \myint d^2\th\, \la X q \bar{q} + \hc
\eal\eeq
Even though $X$ is a background field, we must include a
``kinetic'' term for $X$ to account for the anomalous dimension of
operators that depend on $X$.
(This operator is just the contribution to the cosmological constant.)

This theory has a single classical flat direction with
$\avg{q} = \avg{\bar{q}}$.
We want to compute the effective potential for
$\avg{q} = \avg{\bar{q}} \gg \avg{X}$.
In this case, the largest threshold in the theory is at the
scale
\beq
M_1 = g(M_1) |\avg{q}|,
\eeq
where $g$ is the $\U1$ gauge coupling.
At this scale, the $\U1$ gauge group is completely broken.
The fields that are light below this scale are
$Q$, $\bar{Q}$, and the flat direction $q = \bar{q}$, parameterized
by a field $Y$ defined as
\beq
q = \avg{q} + Y,
\qquad
\bar{q} = \avg{\bar{q}} + Y.
\eeq
The background field $X$ is also present in the low-energy theory.
The effective lagrangian below the scale $M_1$ is therefore
\beq\bal
\scr{L}' &= \myint d^4\th\, \left[
\scr{Z}'_{\vphantom{X}Q} \left( Q^\dagger e^V Q + \bar{Q}^\dagger e^{-V}
\bar{Q}
\right)
+ \scr{Z}'_X X^\dagger X
+ \scr{Z}'_Y Y^\dagger Y
\right]
\\
&\qquad +\, \myint d^2\th\, X Q \bar{Q} + \hc + \cdots,
\eal\eeq
where the elipses denote higher-dimension operators.

The next threshold of interest is the messenger threshold at $M=\la
\avg{X}$.
Below this scale, the effective lagrangian contains only the fields $X$
and $Y$, and it is given by
\beq
\scr{L} = \myint d^4\th\, \left[
\scr{Z}_X X^\dagger X
+ \scr{Z}_Y Y^\dagger Y
\right] + \cdots.
\eeq
We are interested in the effective potential for $Y$ in this
effective lagrangian.
When we continue the couplings into superspace, there will be contributions
to the effective potential for $Y$ from the $Y$ dependence of $\scr{Z}_X$ as
well as the $X$ dependence of $\scr{Z}_Y$.
The field $Y$ does not have renormalizable interactions below
the scale $M_1$, so $\scr{Z}_Y$ does not depend on $X$ at the
renormalizable level.
The contribution to the effective potential we are interested in is
therefore
\beq
V_{\rm eff}(|Y|) = -|\avg{F_X}|^2 \scr{Z}_X(|Y|).
\eeq

We compute $\scr{Z}_X$ using tree-level matching and 1-loop running.
Using the RG equations
\beq\eql{gammaics}
\mu \frac{d \ln \scr{Z}''_X}{d\mu} = -\frac{N_Q}{4\pi^2} \frac{\la^2}{
\scr{Z}''_X \scr{Z}_{\vphantom{X}Q}^{\prime \prime 2}},
\qquad
\mu \frac{d \ln \scr{Z}'_X}{d\mu} = -\frac{N_Q}{4\pi^2} \frac{\la^2}{
\scr{Z}'_X \scr{Z}_{\vphantom{X}Q}^{\prime 2}},
\eeq
we obtain
\beq
\scr{Z}_X = \scr{Z}''_X(\mu_0)
- \frac{N_Q\la^2}{4\pi^2} \int_{\mu_0}^{M_1} \frac{d\mu}{\mu}\,
\frac{1}{\scr{Z}_{\vphantom{X}Q}^{\prime \prime 2}(\mu)}
- \frac{N_Q\la^2}{4\pi^2} \int_{M_1}^M \frac{d\mu}{\mu}\,
\frac{1}{\scr{Z}_{\vphantom{X}Q}^{\prime 2}(\mu)},
\eeq
where $\mu_0$ is a fixed renormalization scale used to define
the theory.
Note that $\scr{Z}_X$ is independent of renormalization scale.
Since we are interested in the $Y$ dependence, we compute
\beq
\frac{\partial \scr{Z}_X}{\partial\ln|Y|} = \frac{N_Q\la^2}{4\pi^2}
\int_{M_1}^{M} \frac{d\mu}{\mu}\, \frac{1}{\scr{Z}'^2_Q(\mu)}\,
\frac{\partial \ln \scr{Z}'_Q(\mu)}{\partial\ln|Y|}.
\eeq
$\scr{Z}'_Q$ does not run in the effective theory $\scr{L}'$, so
we have $\scr{Z}'_Q(\mu) = \scr{Z}''_Q(M_1)$, which gives
\beq
\frac{\partial \ln \scr{Z}'_Q(\mu)}{\partial\ln|Y|}
= \frac{g^2(M_1)}{8\pi^2}.
\eeq
In this way, we obtain
\beq
|Y| \frac{\partial V_{\rm eff}}{\partial|Y|}
= \frac{N_Q\la^2 |\avg{F_X}|^2}{(4\pi^2)^2}
\frac{g^2(M_1)}{\scr{Z}_{\vphantom{X}Q}^2(M_1)}
\ln \frac{M_1}{M}.
\eeq
Note that $M_1$ depends on $|Y|$, so this result automatically gives
the RG-improved form of the effective potential.

\subsection{(S)axion Potential}
There are a number of models for physics beyond the standard model that
involve the spontaneous breaking of a global symmetry at large energy
scales.
For example, ``invisible'' axion models invoke
the breaking of a global $\U1_{\rm PQ}$ symmetry at scales
$10^{10}$--$10^{12}\GeV$ in order to solve the strong \CP problem.
Other global symmetries that may be spontaneously broken include
lepton number and flavor symmetries.

The breaking of a global symmetry will give rise to a massless
Nambu--Goldstone boson (NGB) for every broken generator.
If the global symmetry is broken at a scale where SUSY is
(approximately) unbroken in the visible sector,
then the light bosons must form complete chiral supermultiplets.
There are therefore extra scalars whose mass is protected by SUSY.%
\footnote{If a non-abelian symmetry is broken, some of the Nambu--Goldstone
bosons can belong to the same chiral supermultiplet,
but it can be shown that there are always some ``extra'' scalars.}
We call these fields SNGB's.
The SNGB fields parameterize non-compact directions in the vacuum manifold
in the limit where SUSY is exact,
and different points along the flat direction correpond to different
values for the scale at which the global symmetry is broken.
The SNGB fields will acquire a potential after SUSY breaking, which
determines the \vevs along the flat direction.

As an example, we consider an axion model with colored fields
$R$ and $\bar{R}$ whose mass is determined by the \vev of a field
$\Phi$.
If we write
\beq
\Phi = \avg{\Phi} + A,
\eeq
the imaginary part of $A$ is the axion, while the real part is the SNGB.
The lagrangian is
\beq\eql{axionint}
\bal
\scr{L}'' &= \myint d^4\th\, \scr{Z}_R \left(
R^\dagger e^{V^{(R)}} R
+ \bar{R}^\dagger e^{V^{(\bar{R})}} \bar{R} \right)
\\
&\quad + \myint d^2\th\, \ka \Phi R \bar{R} + \hc
+ \cdots,
\eal\eeq
where we have omitted the messenger sector and standard-model
fields, see \Eq{gamedla}.
The fields $R, \bar{R}$ therefore have a mass
\beq
\mr = \frac{\ka \avg{\Phi}}{\scr{Z}_R(\mr)}.
\eeq
Below this scale, the effective lagrangian $\scr{L'}$
is simply that of ordinary GMSB
together with a kinetic term for the field $\Phi$ (see \Eq{gamedla}).
Below the messenger threshold $M$ the effective lagrangian
$\scr{L}$ is that of the standard model together with kinetic terms
for the singlets $\Phi$ and $X$.
The wavefunction parameter $\scr{Z}_X$ in this effective lagrangian
depends on the $R$ mass, and this contains the leading contribution
to the effective potential for the saxion field.

We can compute $\scr{Z}_X$ using 1-loop running and tree-level matching:
\beq
\scr{Z}_X = \scr{Z}''_X(\mu_0)
+ \int_{\mu_0}^{\mr} \frac{d\mu}{\mu}\, \scr{Z}''_X(\mu) \ga''_X(\mu)
+ \int_{\mr}^{M}\frac{d\mu}{\mu}\, \scr{Z}'_X(\mu) \ga'_X(\mu),
\eeq
where $\ga_X$ is the anomalous dimension of $X$ as defined in \Eq{gammaics}.
The parameter $\scr{Z}_X$ does not run in this effective theory,
so we need not specify a renormalization scale for it.
We compute
\beq
\frac{\partial \scr{Z}_X}{\partial \ln|\Phi|}
&= \int_{\mr}^{M} \frac{d\mu}{\mu}\,
\frac{\partial}{\partial \ln|\Phi|}
\left[ \scr{Z}'_X(\mu) \ga'_X(\mu) \right]
\nonumber\\
&= \frac{N_Q |\la|^2}{4\pi^2} \int_{\mr}^{M} \frac{d\mu}{\mu}\,
\frac{1}{\scr{Z}'^2_Q(\mu)}\,
\frac{\partial \ln \scr{Z}'_Q(\mu)}{\partial \ln|\Phi|}.
\eeq
The \rhs is evaluated using
\beq
\frac{\partial \ln \scr{Z}'_Q}{\partial \ln|\Phi|}
&= \int_{\mr}^{\mu} \frac{d\mu'}{\mu'}\,
\frac{\partial \ga'_Q(\mu')}{\partial\ln|\Phi|}
\nonumber\\
&= -\frac{C_Q}{4\pi^2} \int_{\mr}^{\mu} \frac{d\mu'}{\mu'}\,
g'^4(\mu') \frac{\partial}{\partial \ln|\Phi|}
\left( \frac{1}{g'^2(\mu')} \right)
\nonumber\\
&= \frac{C_Q T_r}{(4\pi^2)^2}
\int_{\mr}^{\mu} \frac{d\mu'}{\mu'}\, g'^4(\mu'),
\eeq
which gives
\beq
|\Phi| \frac{\partial V_{\rm eff}}{\partial |\Phi|} =
-\frac{T_r C_Q N_Q}{(4\pi^2)^3}\,
|\avg{F_X}|^2
\int_{M}^{\mr} \frac{d\mu}{\mu}\,
\frac{1}{\scr{Z}'^2_Q(\mu)}
\int_{\mu}^{\mr} \frac{d\mu'}{\mu'}\, g'^4(\mu').
\eeq
As before, this gives the RG-improved form for the effective potential.
Note that the slope of the potential is negative, indicating that the
saxion \vev is driven away from the origin.

In the opposite limit $\mr \ll M$, it is easy to see that the potential
also decreases as a function of $\mr$.
In the effective theory at the scale $M$, $R$ and $\bar{R}$ get a
positive soft mass-squared from GMSB while $\Phi$ has zero soft mass.
However, the Yukawa coupling $\kappa \Phi R \bar{R}$
drives the $\Phi$ soft mass$^2$ negative in running between $M$
and the scale $\mr$, where $R$ and $\bar{R}$ are integrated out.
(This contribution is analogous to the negative contribution to the
Higgs mass-squared from the top Yukawa coupling.)

Thus in all regions, the potential prefers to push the saxion
\vev, and hence the axion decay constant, to larger values.
Therefore new interactions are needed between the axion and
GMSB sectors in order to stabilize the axion decay constant in the
cosmological and astrophysically desirable window
between $10^{10}$--$10^{12}$~GeV.

\section{Conclusions}
In this paper, we have shown that the renormalization of soft SUSY-breaking
terms is completely determined by the renormalization of SUSY-preserving
terms if the regulator is \susc.
This allows us to calculate certain SUSY-breaking effects in gauge-mediated
theories by performing a \susc calculation and ``analytically continuing'' the
result into superspace.
The method is very powerful, and allows the calculation of interesting
effects at 3-loop order and higher by purely algebraic manipulations.

The formal results that justify these calculations are easy to state
in superspace if the soft SUSY breaking terms are parameterized by
$\th$-dependent terms in the \susc couplings.
If the theory is regulated in a \susc manner, then SUSY is formally preserved
if we regard the bare couplings as superfield spurions.
Our result is that there is a definition of the \emph{renormalized}
couplings that can be similarly grouped into supermultiplets.
Specifically, the renormalized couplings $K_R$ are related to the bare
couplings $K_0$ via a superfield relation of the form
\beq
K_R(\mu) = f(K_0, \La, \mu).
\eeq
The function $f$ determines the renormalization of the \susc couplings
as well as the soft SUSY breaking terms, and is the basis for the
analytic continuation into superspace.
An analogous relation holds between the (renormalized) couplings of
an effective theory and the couplings in a more fundamental theory.

This leads naturally to a definition of the renormalized gauge coupling
chiral superfield
\beq
S(\mu) = \frac{1}{2 g_h^2(\mu)} - \frac{i\Th}{16\pi^2}
- \th^2 \frac{m_{\la,h}(\mu)}{g_h^2(\mu)}
\eeq
as a holomorphic object that is renormalized only at one loop
(to all orders in perturbation theory).
However, the subtraction that defines $S(\mu)$ is not invariant under
constant rescaling of the fields, so the components of $S(\mu)$ do not
correspond directly the usual renormalized couplings.
The real superfield
\beq\eql{final}
R = S + S^\dagger - \frac{T_G}{16\pi^2} \ln(S + S^\dagger)
- \sum_r \frac{T_r}{16\pi^2} \ln \scr{Z}_r
+ \O((S + S^\dagger)^{-1}).
\eeq
is invariant under the field rescaling.
(Here, $r$ runs over the matter representations of the gauge group
$G$ and $T_r$ is the index of $r$;
$\scr{Z}_r$ is the wavefunction factor for the fields in the
representation $r$.)
We show that the lowest components of $R$
\beq
\frac{1}{g^2(\mu)} = \left. R(\mu) \right|,
\qquad
\frac{m_\la(\mu)}{g^2(\mu)} = \left. R(\mu) \right|_{\th^2},
\eeq
are precisely the 1PI gauge
coupling and gaugino mass defined by Euclidean subtraction or by
minimal subtraction in dimensional reduction.
The $\O((S + S^\dagger)^{-1})$ corrections account
for possible scheme dependence in the definition of $R$.
\Eq{final} and much of the story leading up to it
is very similar to the results of \Refs{Russians}, but we emphasize
that all quantities are finite renormalized quantities, and no
reference is made to the Wilsonian renormalization group.

The $\theta^2\bar\theta^2$ component $R$ is given at lowest order by
\beq\eql{strtwo}
\left. R\right|_{\th^2\bar{\th}^2}
= {1\over 8\pi^2} \left[
-T_G m_{\la}^2 + \sum_r T_r m_r^2 \right].
\eeq
and governs the RG evolution of dimension-2 soft terms.
In dimensional reduction $\left. R\right|_{\th^2\bar{\th}^2}$
corresponds to a $1/\ep$ counterterm for the $\ep$-scalar mass.
$\left. R\right|_{\th^2\bar{\th}^2}$ can also be given a 1PI
interpretation: it corresponds to a non-local
$1/p^2$
correction to the propagator
of the gauge supermultiplet.
In the context of dimensional reduction and (modified)
minimal subtraction, our results imply that the simple
extension $1/g^2(\mu) \to R(\mu)$
automatically picks out the so-called $\drbarp$ scheme.

In practice, this result allows one to simply compute the SUSY
breaking components of $R$ (for example) by computing the lowest
component as a function of the \susc bare couplings
(or couplings in an underlying renormalized theory).
This is a \susc calculation, but taking $\th$-dependent
components of the result determines the low-energy
SUSY breaking parameters. For instance,
we have shown that the 2-loop RG
equations for soft terms in $\drbarp$ are directly derived from
the supersymmetric $\beta$-functions and anomalous dimensions.

More remarkably, this approach can be used to
relate leading-log effects computed using
the renormalization group to finite effects, since the result of
taking higher $\th$ components of a logarithm gives effects that are
not logarithmically enhanced:
\beq
\left. \frac{1}{16\pi^2} \ln M \right|_{\th^2\bar{\th}^2}
= \frac{1}{16\pi^2} \frac{\left. M \right|_{\th^2\bar{\th}^2}}
{\left. M \right|}.
\eeq
In this way, we can obtain finite SUSY breaking effects at high loop
order from simple algebraic calculations.
Models with low-energy supersymmetry
breaking mediated by perturbative interactions are  the natural
arena to apply our method.
Indeed, it is precisely in these theories
that it makes more sense to worry also about subleading RG evolution:
this is because the boundary conditions for soft terms are in principle
calculable with comparable accuracy.

Our technique was used to compute a variety of effects at 2-loop
order and beyond.
We computed for the first time the complete subleading corrections
to the gaugino masses (2-loop) and scalar masses (3-loop) in
gauge-mediated models;
we showed how to compute the effective potential for SUSY
flat directions lifted by gauge mediation (2- and 3-loop).
We also proved that gaugino masses are screened from higher-loop
corrections involving couplings in the messenger sector.
Therefore, in the standard gauge mediated scenario,
gaugino masses are rather insensitive on
details of the model. Moreover,
this result also shows that if the gaugino masses are
not generated at one loop (as in the standard case) they will be
generated only from the light matter fields, and will generally
be too light.
This shows that gauge mediation is the unique
way to generate scalar and gaugino masses of the same order through
loop effects.

\def\ijmp#1#2#3{{\it Int. Jour. Mod. Phys. }{\bf #1~}(19#2)~#3}
\def\pl#1#2#3{{\it Phys. Lett. }{\bf B#1~}(19#2)~#3}
\def\zp#1#2#3{{\it Z. Phys. }{\bf C#1~}(19#2)~#3}
\def\prl#1#2#3{{\it Phys. Rev. Lett. }{\bf #1~}(19#2)~#3}
\def\rmp#1#2#3{{\it Rev. Mod. Phys. }{\bf #1~}(19#2)~#3}
\def\prep#1#2#3{{\it Phys. Rep. }{\bf #1~}(19#2)~#3}
\def\pr#1#2#3{{\it Phys. Rev. }{\bf D#1~}(19#2)~#3}
\def\np#1#2#3{{\it Nucl. Phys. }{\bf B#1~}(19#2)~#3}
\def\mpl#1#2#3{{\it Mod. Phys. Lett. }{\bf #1~}(19#2)~#3}
\def\arnps#1#2#3{{\it Annu. Rev. Nucl. Part. Sci. }{\bf #1~}(19#2)~#3}
\def\sjnp#1#2#3{{\it Sov. J. Nucl. Phys. }{\bf #1~}(19#2)~#3}
\def\jetp#1#2#3{{\it JETP Lett. }{\bf #1~}(19#2)~#3}
\def\app#1#2#3{{\it Acta Phys. Polon. }{\bf #1~}(19#2)~#3}
\def\rnc#1#2#3{{\it Riv. Nuovo Cim. }{\bf #1~}(19#2)~#3}
\def\ap#1#2#3{{\it Ann. Phys. }{\bf #1~}(19#2)~#3}
\def\ptp#1#2#3{{\it Prog. Theor. Phys. }{\bf #1~}(19#2)~#3}

\end{document}